\documentclass[prd, 10pt,aps,nofootinbib, floatfix, notitlepage,twocolumn]{revtex4}

\usepackage{amsmath,amsfonts,amssymb}
\usepackage{mathrsfs}
\usepackage{graphicx}
\usepackage[english]{babel} 
\usepackage{hyperref} 
\usepackage{url} 
\usepackage{color}
\usepackage{bbold}
\usepackage{adjustbox}
 \usepackage[utf8]{inputenc} 

\newcommand{\be}{\begin{equation}}
\newcommand{\ee}{\end{equation}}
\newcommand{\bes}{\begin{equation*}}
\newcommand{\ees}{\end{equation*}}

\begin{document}

\title{Cosmological implications of ultra-light axion-like fields}
\author{Vivian Poulin$^1$}
\author{Tristan L.~Smith$^2$}
\author{Daniel Grin$^3$}
\author{Tanvi Karwal$^1$}
\author{Marc Kamionkowski$^1$}
\affiliation{$^1$Department of Physics and Astronomy, Johns
				Hopkins University, 3400 N.\ Charles St., Baltimore, MD 21218, United States}
\affiliation{$^2$Department of Physics and Astronomy, Swarthmore College, 500 College Ave., Swarthmore, PA 19081, United States}
\affiliation{$^3$Department of Physics and Astronomy, Haverford College, 370 Lancaster Avenue, Haverford, PA 19041, United States}

\begin{abstract}
Cosmological observations are used to test for imprints of an ultra-light axion-like field (ULA), with a range of potentials $V(\phi)\propto[1-\cos(\phi/f)]^n$ set by the axion-field value $\phi$ and decay constant $f$. Scalar field dynamics dictate that the field is initially frozen and then begins to oscillate around its minimum when the Hubble parameter drops below some critical value.  For $n\!=\!1$, once dynamical, the axion energy density dilutes as matter; for $n\!=\!2$ it dilutes as radiation and for $n\!=\!3$ it dilutes faster than radiation.  Both the homogeneous evolution of the ULA and the dynamics of its linear perturbations are included, using an effective fluid approximation generalized from the usual $n=1$ case. ULA models are parameterized by the redshift $z_c$ when the field becomes dynamical, the fractional energy density $f_{z_c} \equiv \Omega_a(z_c)/\Omega_{\rm tot}(z_c)$ in the axion field at $z_c$,  and the effective sound speed $c_s^2$. Using {\em Planck}, BAO and JLA data, constraints on $f_{z_c}$ are obtained. ULAs are degenerate with dark energy for all three potentials if $1+z_c \lesssim 10$.  When $3\times10^4 \gtrsim 1+z_c \gtrsim 10 $, $f_{z_c}$ is constrained to be $ \lesssim 0.004 $ for $n=1$ and $f_{z_c} \lesssim 0.02 $ for the other two potentials. The constraints then relax with increasing $z_c$. These results strongly constrain ULAs as a resolution to cosmological tensions, such as discrepant measurements of the Hubble constant, or the EDGES measurement of the global 21 cm signal.
\end{abstract}

\date{\today}

\maketitle

\section{Introduction}
The nature of the dark matter (DM) and dark energy (DE) that dominate our universe today is one of the biggest mysteries of modern cosmology. The dominant paradigm is the $\Lambda$CDM model, in which DM is a cold, gravitationally interacting particle,  while DE is a pure cosmological constant. Remarkably, this simple model is consistent with precise measurements of the cosmic microwave background (CMB) anisotropies by {\em Planck} \citep{Ade:2015xua}, but remains purely parametric. 

Ultra light axion-like (ULA) fields arise generically in string theory \citep{Arvanitaki:2009fg,Marsh:2015xka}. They may be cosmologically relevant, contributing to the cold Dark Matter (CDM) \textit{and} DE in our universe (see \citep{Marsh:2015xka} and references therein).  These models have also been invoked to solve tensions within the $\Lambda$CDM model, calling on the presence of an early dark energy (EDE) phase \citep{Karwal:2016vyq,Hill:2018lfx}. 
 
For example, increasingly precise measurements of the local expansion rate have led to a potentially significant disagreement (see, e.g., Ref.~\citep{Feeney:2017sgx}) between measurements of the Hubble constant inferred from the CMB \citep{Ade:2015xua} at high redshifts and Cepheid variables/supernovae at low redshifts \citep{Riess:2016jrr}. Additionally, if the recently claimed measurement of $21$-cm absorption at $z \sim 20$ by the EDGES experiment \citep{Bowman:2018yin} withstands experimental scrutiny \citep{Hills:2018vyr}, the presence of such early cosmological structure \citep{Bozek:2014uqa} sets a lower bound on the ULA mass of $\sim 10^{-21}~{\rm eV}$ \citep{Lidz:2018fqo}, if ULAs compose all of the dark matter.
 
The apparent anomalously low baryon temperature measured by EDGES could indicate that the expansion history at high redshifts could differ from standard assumptions. These observations could be explained through the cosmological effects of a collection of scalar fields, as envisioned in the `string-axiverse' scenario \citep{Svrcek:2006yi,Arvanitaki:2009fg,Stott:2017hvl}. These fields would also affect a variety of cosmological observables, such as CMB and matter power-spectra \citep{Hlozek:2014lca,Mukherjee:2018oeb}. 

In this paper we explore the observational implications of a cosmological scalar field with a potential of the form $V_n(\phi) \propto [1-\cos(\phi/f)]^{n}$ that becomes dynamical at a range of times, which arises non-perturbatively and  breaks the approximate ULA shift symmetry.
The standard axion potential is obtained in the $n=1$ case, while higher-$n$ potentials may be generated by higher-order instanton corrections \citep{Kappl:2015esy}.

Here $\phi$ denotes the field value and $f$ the ULA decay constant. These fields become dynamical as the Hubble parameter decreases, eventually settling down at the minima of their potentials. Up to the point when the fields become dynamical (i.e.~during the period of `slow-roll' evolution) their equations of state are dark-energy like: $w_a\simeq -1$.  

Soon after the field becomes dynamical it starts to oscillate and, when averaged over the oscillation period, has an equation of state equal to $w_a\simeq (n-1)/(n+1)$ for a potential of the form $V_n(\phi) \propto \phi^{2n}$ \citep{PhysRevD.28.1243}.  As the field oscillates, its energy density dilutes as cold dark matter (CDM) for $n=1$, for $n=2$ it dilutes as radiation and for $n=3$ it dilutes faster than radiation. With a statistical ensemble of such fields (i.e. the `axiverse') the universe may have gone through several periods of `anomalous' expansion, alleviating the coincidence problem today \citep{Griest:2002cu,Linder:2010wp,Kamionkowski:2014zda,Emami:2016mrt,Karwal:2016vyq}, and possibly reducing the Hubble constant tension \citep{Karwal:2016vyq} and explaining the anomalously low baryon temperature inferred by the EDGES experiment \citep{Hill:2018lfx}. This general scenario may also provide a way to connect the physics of cosmic inflation to our current period of accelerated expansion \citep{Kamionkowski:2014zda}.  

Here, we extend previous work in several significant ways. First, we present a fluid approximation that parameterizes the ULA dynamics for arbitrary $n$ in terms of the redshift when the field becomes dynamical, $z_c$, and the fractional energy density in the axion field at $z_c$, $f_{z_c} \equiv \Omega_a(z_c)/\Omega_{\rm tot}(z_c)$. A key result of this work is the inclusion of ULA perturbations using an effective fluid approach for $n=2$ and $n=3$.  These perturbations can be approximately described by a time-averaged fluid component with a time and scale dependent effective sound speed \citep{Hu:2000ke,Hwang:2009js,Marsh:2010wq,Park:2012ru,Hlozek:2014lca,Marsh:2015xka,Noh:2017sdj} within the `generalized dark matter' parameterization \citep{Hu:1998kj}.

Past applications of this effective fluid approach were restricted to a scalar field of mass $m$ in a quadratic potential. The effect of anharmonicities on the background has been explored (e.g. Ref.~\citep{Lyth:1991ub}), and in Ref.~\citep{Cembranos:2015oya}, a preliminary effective fluid treatment of anharmonic scalar fields was considered. Similar results are obtained by taking the Schr\"{o}dinger limit of the Klein-Gordon equation for small length scales, as shown in Ref.~\citep{Fan:2016rda}. Here, we generalize past work systematically to anharmonic potentials ($n=2$, and $3$), deriving a new straightforward expression for the sound speed $c_{\rm eff}$ which is easy to compute once the behavior of the homogeneous field is known. Moreover, we derive a mapping between this parametrization and the ULA mass, decay constant and initial field value. We show that our fluid formalism is adequate for $n\leq 3$, but breaks down for larger values of $n$ for which the period of oscillation is never much shorter than a Hubble time.

Using {\em Planck}, measurements of the baryon acoustic oscillations (BAO) and the Joint Light-Curve Analysis (JLA) data \citep{Betoule:2014frx}, we place constraints on ULAs in the $n=1,2$ and 3 models.
Using a Markov Chain Monte Carlo (MCMC) analysis, we are able to fully explore degeneracies between the ULA parameters and the standard cosmological parameters. We derive constraints on $f_{z_c}$ as a function of $z_c$. We find in particular that $f_{z_c}$ becomes partially degenerate with dark energy for all three potentials once $1+z_c > 10$.  When $3\times10^{4} \lesssim 1+z_c \lesssim 10$, we find that $f_{z_c}$ is constrained to be $ \lesssim 0.004 $ for matter-dilution and $f_{z_c} \lesssim 0.02 $ for the other two potentials. The constraints then relax with increasing $z_c$, but we demonstrate that current measurements of the CMB\footnote{Naturally, alternative probes such as BBN can constrain the parameter space further at such times.} require that $f_{z_c}$ be less than unity as early as $z_c = 10^{10}$. Remarkably, we find that the details of the ULA dynamics could distinguish its effects from other cosmological components, even if the ULA time-averaged equation of state is equal to zero (CDM-like) or 1/3 (radiation-like). 

The organization of this paper is as follows. In Sec.~\ref{sec:cosmo_ULAs}, we review the basics of the cosmological dynamics of ULAs by laying out the equations for the homogeneous field dynamics and introducing the dynamics of the perturbed field. We also present our fluid approximation and how it maps to the ULA theory parameters.  Equipped with this formalism, we describe in Sec.~\ref{sec:perturb} the rich dynamics of ULA perturbations. Then, in Sec.~\ref{sec:observables} we  calculate the CMB and matter power-spectra that arise in our scenario using a modified version of the \texttt{CLASS} Boltzmann code\footnote{\url{http://class-code.net}} \citep{Lesgourgues:2011re,Blas:2011rf,Lesgourgues:2011rg,Lesgourgues:2011rh}. In Sec.~\ref{sec:constraints}, we use the \textsc{MontePython}\footnote{\url{http://baudren.github.io/montepython.html}} \citep{Audren:2012wb} MCMC package to obtain constraints on our scenario. We discuss implications for cosmological tensions in Sec.~\ref{sec:implications_for_tensions}. We conclude in Sec.~\ref{sec:conclusions}. In Appendix~\ref{sec:app_derivs}, we obtain the generalized effective fluid equations for anharmonic potentials and the effective sound speed for arbitrary $n$, a result which may be of interest beyond the specific ULA scenario considered here. We compare our fluid formalism to exact solutions of the Klein-Gordon (KG) equations in Appendix~\ref{sec:app_full_vs_approx}. 

\section{The cosmological dynamics of ULAs}
\label{sec:cosmo_ULAs}
\subsection{Background dynamics}

The background dynamics of a ULA have a simple description. The field is initially pinned at some value due to Hubble friction. Once the expansion rate drops below some critical value (related to the mass of the ULA), the field is free to evolve to the minimum of the potential. It then oscillates around the bottom of its potential such that its energy density is diluted due to the subsequent expansion. 

The homogeneous Klein-Gordon (KG) equation of motion for the field is given by
\begin{equation}
   \ddot{\phi} + 3 H \dot{\phi} + \frac{dV_n(\phi)}{d\phi} = 0.
\end{equation}
 The ULA potential is given by
\begin{equation}
    V_n(\phi) = \Lambda^4(1-\cos \phi/f)^n,
\end{equation} 
where $f$ is the energy scale at which the global $U(1)$ related to axions is spontaneously broken. The ULA homogeneous energy-density and pressure are 
\begin{eqnarray}
    \rho_a &=& \frac{1}{2} \dot{\phi}^2 + V_n(\phi),\\
     P_a &=& \frac{1}{2} \dot{\phi}^2 - V_n(\phi).
\end{eqnarray}
The Hubble equation can be written 
\begin{equation}
    H=H_0 E(a) = H_0 \sqrt{\Omega_m(a) + \Omega_r(a) + \Omega_\Lambda + \Omega_a(a)},
\end{equation}
where $\Omega_X \equiv \rho_X/\rho_{\rm crit}$ and $\rho_{\rm crit} = 3H_0^2 M_P^2$, where $M_P \equiv (8\pi G)^{-1/2}$ is the reduced Planck mass. In order to solve these equations numerically it is useful to redefine the variables so that they are dimensionless. If we define $\Theta \equiv \phi/f$, $m\equiv \Lambda^2/f$, $\alpha \equiv f/M_P$, $x \equiv H_0 t$, and $\mu \equiv m/H_0$ these equations can be written 
\begin{eqnarray}
V_n(\Theta) &=& \mu^2 \alpha^2(1-\cos \Theta)^n,\label{eq:Vn}\\
\Theta''&=& - 3 E \Theta' - \alpha^{-2} \frac{d V_n}{d\Theta},\label{eq:KG}\\
\Omega_a(a) &=& \frac{1}{3}\left[\frac{1}{2} \alpha^2\Theta'^2 + V_n(\Theta)\right],
\end{eqnarray}
where a prime indicates a derivative with respect to $x$. 

Before the field starts to oscillate it undergoes `slow-roll' evolution (that is, $ \dot{\phi}^2/2 \ll V$ and the dynamics are dominated by Hubble friction) which we will refer to as an `early dark energy' (EDE) phase. To obtain a useful parameterization for all the models under consideration, we have found an analytic approximation to the initial field evolution. First, we expand the potential to linear order around the initial field value $\Theta_i$ to obtain a solution for the field evolution (assuming that $\Theta_i' \rightarrow 0$ as $x \rightarrow 0$):
\begin{eqnarray}
  \Theta(x) &\simeq& \Theta_i+\frac{\sin (\Theta_i) \left(\, _0F_1\left[\frac{1}{2} (3 p+1);\mathcal{A} x^2\right]-1\right)}{n \cos \Theta_i+n-1}, \label{eq:ThetaApprox} \\
  &\simeq& \Theta_i-\frac{\mu^2 n x^2 \sin \Theta_i (1-\cos \Theta_i)^{n-1}}{2 (3 p+1)} + \mathcal{O}(\mathcal{A}^2 x^4) \nonumber
\end{eqnarray}
where $_0F_1$ is the confluent hyper-geometric function and
\begin{equation}
    \mathcal{A} \equiv \frac{1}{4} \mu^2 n  (1-\cos \Theta_i)^{n-1} (1-n\cos \Theta_i-n),
\end{equation}
and where $ \Theta_i$ is the initial value of the field at $x=0$ and $a'/a = p/x$ so that during radiation domination $p=1/2$ and during matter domination $p=2/3$. When numerically solving for the evolution of the homogeneous scalar field, we take the initial field value to be $0<\Theta_i < \pi$ and the initial velocity of the field is determined by the curvature of the potential at $\Theta_i$ through Eq.~(\ref{eq:ThetaApprox}). We set $p=1/2$ since the field is always initialized during radiation domination.

\begin{figure}
    \centering
    \includegraphics[scale=0.33]{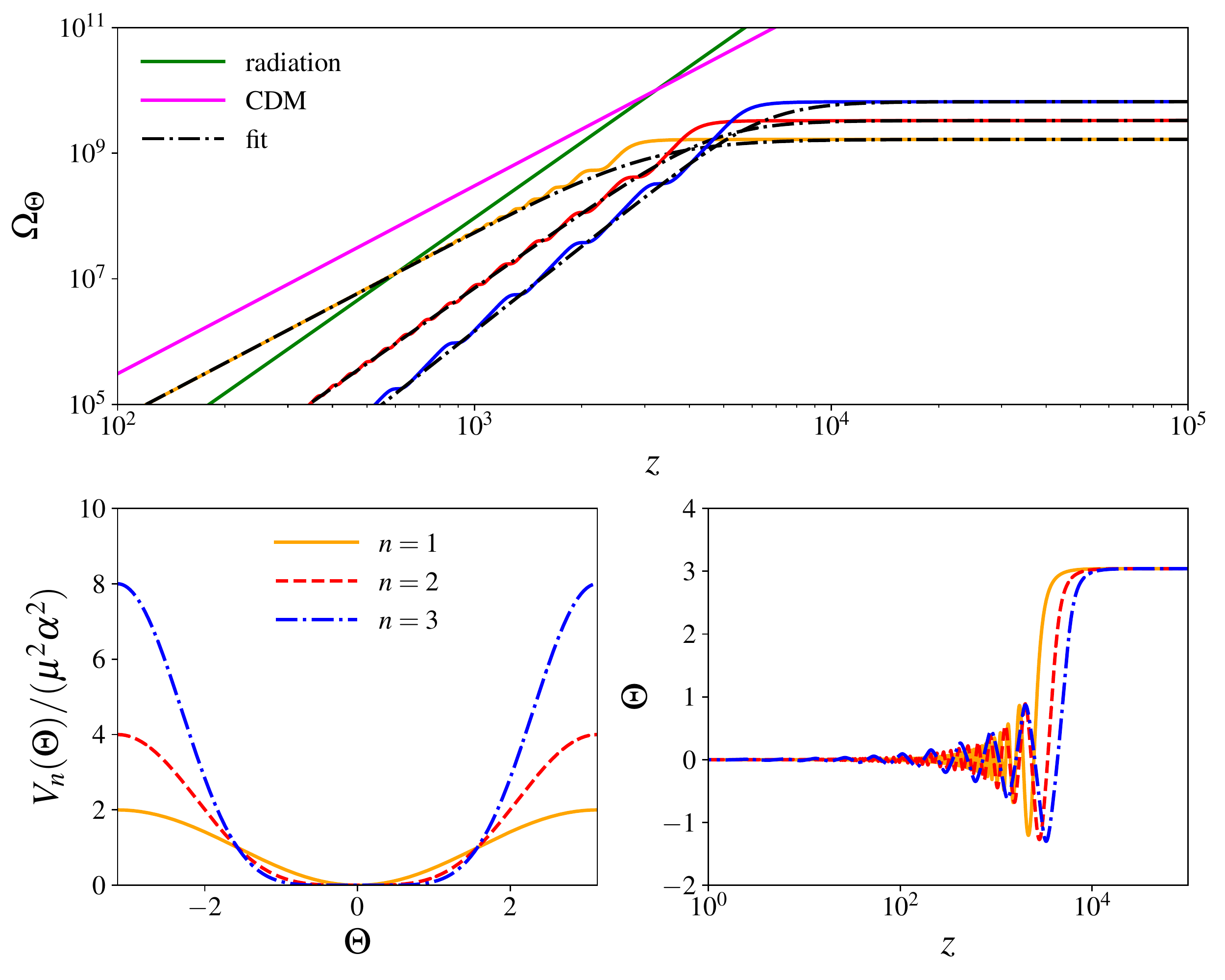}
    \caption{The evolution of the background field with $\mu = 10^6$, $\alpha = 0.05$, and $\Theta_i = \pi-0.1$ for the three forms of the axion potential explored in this paper.}
    \label{fig:background_evo}
\end{figure}

After a period of slow-roll evolution, the field transitions to an oscillatory phase with a decreasing amplitude due to the dilution of the field's energy density from expansion. The potential during the oscillating phase takes the form $V_n(\Theta) \simeq 2^{-n} \mu^2 \alpha^2 \Theta^{2n}$ so that for $n=1$ the field undergoes simple harmonic oscillation with a frequency which is independent of its amplitude and for $n>1$ the oscillations are anharmonic and the frequency depends on the amplitude. We show the evolution of $\Theta$ for the three forms of the potential considered here in Fig.~\ref{fig:background_evo}. 

Once oscillating, over time-scales shorter than a Hubble time the field evolves according to the equation of motion 
\begin{equation}
    \Theta '' + \alpha^{-2} \frac{dV_n}{d\Theta} = 0.
\end{equation}
Furthermore if we assume that the oscillation frequency $\varpi \gg H$, the total energy will be approximately conserved over several oscillations so that we have 
\begin{equation}
    \frac{1}{2} \Theta'^2 +\alpha^{-2}V_n(\Theta) = \alpha^{-2}V_n(\Theta_m),\label{eq:energy}
\end{equation}
where $\Theta_m$ is the maximum field value, reached when $\Theta' = 0$. 
We can use the virial theorem to write $\langle 1/2 \Theta '^2 \rangle = n \alpha^{-2} \langle V_n \rangle$ so that
\begin{eqnarray}
    \langle \Omega_\Theta \rangle &\simeq& \frac{1}{3}\frac{\mu^2\alpha^2}{2^n}\Theta_m^{2n} \simeq \Omega_{a,0}a^{-3(1+w_n)},
\end{eqnarray}
which shows that, due to the expansion of the universe, the maximum field value will decrease as $\Theta_m \propto a^{-3/(1 + n)}$. As in Ref.~ \citep{PhysRevD.28.1243}, we find that the axion energy density is constant at early times and decays as $a^{-3(1+w_n)}$ with
\begin{equation}
    w_n \equiv \frac{n-1}{n+1}.\label{eq:wn}
\end{equation}
With this we will parameterize the axion energy density by 
\begin{equation}
 \Omega_a(z)= \frac{2 \Omega _{a}(z_c)}{\left[(1+z_c)/(1+z)\right]^{3( w_n+1)}+1},\label{eq:omegaFit}
 \end{equation}
 which has an associated equation of state 
 \begin{equation}
     w_a(z) = \frac{1+w_n}{1+[(1+z)/(1+z_c)]^{3(1+w_n)}}-1,
     \label{eq:wphi}
 \end{equation}
 and which asymptotically approaches $-1$ as $a \rightarrow 0$ and $w_n$ for $z \ll z_c$. We show a comparison between the exact axion energy density and our parameterization in Fig.~\ref{fig:background_evo}. This shows that when $n=1$, the homogeneous axion energy density dilutes like matter once the field is dynamical. On the other hand it dilutes like radiation when $n=2$. When $n\geq 3$, dilution is faster than radiation \footnote{A qualitatively similar stiff dilution phase occurs in \textit{complex} scalar-field dark-matter models, with the relevant phenomenology discussed in Refs. \citep{Li:2013nal,Li:2016mmc}.}.

\subsection{On the validity of the fluid approximation}
From Eq.~(\ref{eq:energy}) we can compute the time for one full oscillation:
\begin{eqnarray}
    T &\simeq& 4 H^{-1}_0 \int_0^{\Theta_m} \frac{d\Theta}{\sqrt{2\alpha^{-2}[V(\Theta_0)-V(\Theta)]}},\\
    &=& 4 H_0\frac{\sqrt{\pi } 2^{\frac{n-1}{2}} \Theta_m^{1-n} \Gamma \left(1+\frac{1}{2 n}\right)}{\mu  \Gamma \left(\frac{n+1}{2 n}\right)}
    \label{eq:perdiod}
\end{eqnarray}
 This leads to an angular frequency 
\begin{eqnarray}
\varpi &=& \varpi_0 a^{-3 w_n}, \label{eq:omegavar}\\
\varpi_0 &=&H_0\frac{\sqrt{\pi } 2^{-\frac{n^2+1}{2 n}} \Omega_{\Theta, 0}^{\frac{n-1}{2 n}} \Gamma \left(\frac{n+1}{2 n}\right) (\alpha  \mu )^{1/n}}{\alpha  \Gamma \left(1+\frac{1}{2 n}\right)}. \label{eq:omegan}
\end{eqnarray}
This shows that the angular frequency is only constant if $n=1$; for $n>1$ the oscillation frequency \emph{decreases} in time \citep{Johnson:2008se}.  In particular, the fluid approximation is only accurate if $\varpi/H \gg 1$ and, assuming that the axion field never dominates the energy budget, we have 
 \begin{equation}
     \frac{\varpi}{H} \propto \begin{cases}
      a^{  (5 - n)/(1 + n)}&\  a<a_{\rm eq},\\
       a^{  6/(1 + n)-3/2} &\ a>a_{\rm eq},
\end{cases}
 \end{equation}
 where $a_{\rm eq} \equiv \Omega_{r,0}/\Omega_{m,0}$ is the value of the scale-factor at matter/radiation equality. This ratio increases with time for $n<5$ during  radiation domination and for $n<3$ for matter domination. During a period of accelerated expansion the ratio will \emph{decrease} in time for any positive value of $n$. Therefore, in this work we limit our study to $n\leq3$ such that if $\varpi/H \gtrsim 1$ at the start of the oscillatory phase, then the ratio will remain large up until almost today, when the latest epoch of cosmic acceleration began.

\subsection{Perturbed dynamics in the fluid formalism: a first look}

Linear perturbations to the axion field will develop and evolve according to the perturbed Klein-Gordon equation. However, these equations are computationally expensive to solve and would not allow us to  scan over the parameters of the ULA theory and the standard cosmological parameters. Since the oscillations of the scalar field generally occur with periods much shorter than a Hubble time, much of the dynamics can be captured by averaging over the oscillations and dealing with fluid equations \citep{PhysRevD.28.1243}. The equations governing the evolution of density and bulk velocity perturbations can be written in terms of fluid variables in the synchronous gauge as \citep{Hu:1998kj}
\begin{eqnarray}
    \dot{\delta}_a&=&-(1+w_a)\bigg(\theta_a+\frac{\dot{h}}{2}\bigg)-3(c_s^2-w_a){\cal H} \delta_a\nonumber\\
     & & -9(1+w_a)(c_s^2-c_a^2){\cal H}\frac{\theta_a}{k^2}\,,\\
    \dot{\theta}_a&=&-(1-3c_s^2){\cal H}\theta_a+\frac{c_s^2k^2}{1+w_a}\delta_a\,,
\end{eqnarray}
where in these equations the dot refers to a derivative with respect to conformal time. From the background dynamics, $w_a$ is known. Note that the effective sound speed $c_s^2\equiv \delta p/\delta \rho$, is possibly different from unity for an ULA, and the adiabatic sound speed
\begin{equation}
    c_{\rm a}^2 \equiv \frac{\dot P_a}{\dot \rho_{a}} =  w_a-\frac{\dot{w}_a}{3(1+w_a){\cal H}}\,. \label{eq:cad}
\end{equation}

The adiabatic sound speed is straight forward to calculate since it depends only on background quantities. Using the initial EDE evolution of the field given in Eq.~(\ref{eq:ThetaApprox}) and assuming $\dot{\phi}_i=0$ \footnote{In our model, this is naturally realized because of the large Hubble friction at early times.}, one can show that $c_{\rm a}^2 \simeq -7/3$ \citep{Hlozek:2014lca} during slow-roll for any form of the potential. 

In the approximation for $w_a$ given by Eq.~(\ref{eq:wphi}) the adiabatic sound-speed during the EDE period is given by 
\begin{equation}
    c_{\rm a}^2 = -\frac{3 n+1}{n+1}, \label{eq:cadSR}
\end{equation}
it then evolves to $w_a$ once the field starts oscillating.
At early times, except for the case $n=2$, this parameterized adiabatic sound-speed differs from the exact value of -7/3 (with a range $-7/3 \leq c_{\rm a}^2 \leq -5/2$). We have checked that given that both the exact and parameterized $c_{\rm a}^2$ are negative and of order unity, our parameterization gives a good approximation to the exact evolution of the perturbations.  We show a comparison between the exact mode evolution and the approximate mode evolution in Appendix~\ref{sec:app_full_vs_approx}.

Finally, in order to utilize the GDM equations of motion, we must determine $c_s^2$. During the EDE phase $c_s^2=1$ for a slowly rolling scalar field, but deviates strongly from 1 once the field starts oscillating. We discuss our derivation of the time-averaged effective sound speed in Appendix~\ref{sec:app_derivs}.  We find that for a ULA potential which takes the form $V \propto \phi^{2n}$ around the minimum:
\begin{eqnarray}
 c_s^2 = \frac{2 a^2 (n-1) \varpi ^2+k^2}{2 a^2 (n+1) \varpi ^2+k^2},\label{eq:ceff2}
\end{eqnarray}
with the frequency $\varpi$ given by Eq.~(\ref{eq:omegavar}). 
We discuss the dynamics of perturbations in Sec.~\ref{sec:perturb}. Before entering into these details, we relate our parametrization to the ULA theory parameters.

\subsection{Approximate translation between model and theory parameters}
\label{sec:translation}

The axion model is fully specified by four `theory' parameters: the potential-index $n$, the initial field value $\Theta_i \equiv \phi_i/f$, the mass parameter $\mu \equiv m/H_0$, and the coupling parameter $\alpha \equiv f/M_P$. Our model is also described by four parameters: the redshift $z_c$ when the field begins to oscillate, the energy density of the field $\Omega_\phi(z_c)$ at $z_c$, the time-averaged equation of state $w_n$ during oscillations, and the scale dependence $\varpi_0$ of the effective sound-speed. The equation of state $w_n$ and the index $n$ are related through Eq.~(\ref{eq:wn}) and $\varpi_0$ is related to $\alpha$ and $\mu$ through Eq.~(\ref{eq:omegan}). The last two parameters are related by more involved expressions, as we now discuss. 

First, note that we can use Eq.~(\ref{eq:ThetaApprox}) and Eq.~(\ref{eq:omegaFit}) to relate $\Omega_a(z_c)$ to $\mu$, $\alpha$, and $\Theta_i$ by computing the energy density in the axion field at very early times:
\begin{equation}
   \Omega_a(z_c) = \frac{1}{6} \alpha ^2 \mu^2 (1-\cos \Theta_i)^n.\label{eq:Omac}
\end{equation}
We can obtain an approximate expression for $z_c$ by noting that the field starts to oscillate soon after the field evolves away from its initial value, $\Theta_i$. We can compute the time at which the field starts to evolve using the approximate evolution of $\Theta(x)$ given in Eq.~(\ref{eq:ThetaApprox}).  We define $x_c$ as the time at which the field evolves to some fraction of its initial value, $\Theta(x_c) = \mathcal{F}\Theta_i$:
\begin{equation}
    x_c \equiv \frac{(1-\cos \Theta_i)^{\frac{1-n}{2}} }{\mu }\sqrt{\frac{(1-\mathcal{F}) (6 p+2) \Theta _i}{n \sin \Theta _i}}.\label{eq:xc}
\end{equation}
We can relate this to $z_c$ by using the fact that during radiation or matter domination the Hubble parameter is given by $E \simeq p/x$ so that 
\begin{equation}
    E(z_c) \simeq \frac{p}{x_c},\label{eq:actoxc}
\end{equation}
where, as before, $p=1/2$ for $z_c > z_{\rm eq}= \Omega_{r,0}/\Omega_{M,0} \simeq 10^{-5}$ and $p=2/3$ for $z_c < z_{\rm eq}$. 
We compare the full field evolution by solving Eq.~(\ref{eq:KG}) to our model in Eq.~(\ref{eq:omegaFit}). We find that $z_c$ is most accurately determined when we choose $\mathcal{F} = 7/8$. Note that our approximate solution for the field evolution in Eq.~(\ref{eq:ThetaApprox}) fails in the limit $\Theta_i \rightarrow \pi$ and we have found that this mapping can reproduce the full dynamics up until $\Theta_i \simeq 3$. Also note that for $n=1$ and $\Theta_i \ll 1$ our results give $H(z_c) \simeq m$ which agrees with previous work \citep{Marsh:2010wq}. 

This mapping can be used to go from our model parameters to the theory parameters. Assuming that the field makes up a small fraction of the total energy density at $z_c$, we can use Eq.~(\ref{eq:actoxc}) to determine $x_c$ and then Eq.~(\ref{eq:xc}) provides a relationship between $\mu$ and $\Theta_i$. Given $\Omega_\phi(z_c)$, Eq.~(\ref{eq:Omac}) provides a relationship between $\mu$, $\alpha$, and $\Theta_i$.  Combining these together we can then write $\varpi_0$ as a function of $z_c$, and $\Theta_i$:
\begin{eqnarray}
    \varpi_0(z_c,\Theta_i,n) &=&H_0\mu(z_c,\Theta_i,n)\left(1-\cos \Theta _i\right)^{\frac{n-1}{2}}\mathcal{G}(z_c,n){},\nonumber \\ \label{eq:omega_n}\\
    \mathcal{G}(z_c,n) &\equiv& \frac{\sqrt{\pi }\Gamma \left(\frac{n+1}{2 n}\right)}{\Gamma \left(1+\frac{1}{2 n}\right)} 2^{-\frac{n^2+1}{2 n}} 3^{\frac{1}{2} \left(\frac{1}{n}-1\right)}   \nonumber \\ &\times&(1+z_c)^{\frac{6}{n+1}-3}\left[(1+z_c)^{\frac{-6 n}{n+1}}+1\right]^{\frac{1}{2} \left(\frac{1}{n}-1\right)}.
\end{eqnarray}
This shows that, in principle, the homogeneous and perturbative effects of this field on cosmological observation can give us enough information to reconstruct all of the theory parameters. Said another way, an estimate of $z_c$ and $\Omega_{a}(z_c)$ from the homogeneous dynamics of the field will determine the evolution of perturbations up to the unknown initial field value, $\Theta_i$; an estimate of $\varpi_0$ from the perturbations then determines $\Theta_i$. 

Finally, we can use these expressions to relate the theory parameters to the model parameters. In particular Eqs.~(\ref{eq:xc}) and (\ref{eq:omega_n}) show that $z_c$ and $\varpi_0$ are both determined by $\mu$ and $\Theta_i$. These can then be combined to give an estimate of the fractional contribution of the ULA to the total energy density $f_{z_c}$ at $z_c$.
Recall that these expressions have assumed $f_{z_c} \ll 1$ and our analytic expressions for the field evolution are only accurate for $\Theta_i \lesssim \pi-0.1$. 

\section{Detailed study of ULA perturbations in the fluid approximation}
\subsection{Setup and initial conditions of perturbations}\label{sec:perturb}

As explained previously, we solve for the ULA dynamics using the GDM equations of motion \citep{Hu:1998kj}, which require the specification of the ULA equation-of-state $w_a$, the adiabatic sound speed $c_{\rm a}^2$, and effective sound speed $c_s^2$. During slow roll, generic scalar fields have that $w_a \simeq -1$, $c_{\rm a}^2 \simeq -7/3$, and $c_s^2 =1$. Since $w_a \simeq -1$ the linear perturbation equations written in terms of the velocity perturbation $\theta_a$ are unstable.  To deal with this we solve the evolution of the perturbations in terms of the heat-flux, $u_a \equiv (1+w_a) \theta_a$ \citep{Hlozek:2014lca}. \begin{eqnarray}
    \dot{\delta}_a&=&-\bigg[u_a+(1+w_a)\frac{\dot{h}}{2}\bigg]-3(c_s^2-w_a){\cal H} \delta_a\nonumber\\
     & & -9(c_s^2-c_{\rm a}^2){\cal H}\frac{u_a}{k^2}\,,\\
    \dot{u}_a&=&-(1-3c_s^2){\cal H}u_a+3{\cal H}(w_a-c_{\rm a}^2)u_a\nonumber\\
    & &  +c_s^2k^2\delta_a\,.
\end{eqnarray}
In practice, when $z>z_c$, we set $w_a \simeq -1$, $c_s^2 =1$ and $c_{\rm a}^2$ is given by Eq.~(\ref{eq:cadSR}). 
During the oscillatory phase, when $z < z_c$, $c_s^2$ is given by the time and scale-dependent effective sound speed in Eq.~(\ref{eq:ceff2}), $c_{\rm a}^2$ is given by Eq.~(\ref{eq:cad}) with $w_a$ given by Eq.~(\ref{eq:wphi}). Abrupt changes in these quantities can lead to the generation of transients in numerical solutions. We have verified that these had no significant effects on the predicted power spectra used to constrain this model.  A comparison between the approximate and exact ULA evolution is discussed in Appendix \ref{sec:app_full_vs_approx} and shows very good agreement. 

In general, adiabatic initial conditions on super-Hubble scales are expected when the perturbations within each component are due to a single degree of freedom (e.g. slight time delay in the decay of the inflaton field) and lead to simple relations of the type
\begin{equation}
    \frac{\delta_i(\tau,\vec{x})}{1+w_i} =  \frac{\delta_{i'}(\tau,\vec{x})}{1+w_{i'}}=-\frac{h}{2}\,,
\end{equation}
where $i$ and $i'$ are two species  and $h\sim(k\tau^2)$ corresponds to the growing mode solution of a fourth order linear differential equation for the trace of the metric perturbation in the synchronous gauge \citep{Ma:1995ey}. For a species with zero non-adiabatic sound speed, this would typically be enough. However, a fluid with  $c_s^2\neq c_a^2$ does not generally obey such relations. In the ULA scenario considered here the ULA component is always subdominant on superhorizon scales and at early times. In that case, the ULA perturbations fall inside the gravitational potential wells created by the radiation component, such that there is a generic attractor solution \citep{Ballesteros:2010ks}
\begin{eqnarray}
\delta_a &=& -\frac{C}{2}(1+w_a) \frac{4-3 c_s^2}{4-6 w_a +3 c_s^2} (k\tau)^2, \label{eq:deltaSH}\\
u_a &=& - \frac{C}{2}(1+w_a) \frac{c_s^2}{4-6 w_a + 3 c_s^2}(k \tau)^3 k, \label{eq:thetaSH}
\end{eqnarray}
where $C$ is the initial amplitude and $\tau$ is the conformal time. Note that we take $\delta_a = u_a = 0$ initially since these quantities are quickly driven to the attractor solution \citep{Ballesteros:2010ks}. 

More generally, if the axion symmetry-breaking scale $f>H_{\rm I}$ (where $H_{I}$ is the inflationary Hubble parameter), axions will carry isocurvature perturbations, as a light relic present during the inflationary era (see Ref.~\citep{Hlozek:2017zzf} and references therein). This will change the height of the Sachs-Wolfe plateau and alter the phases of CMB acoustic peaks. Limits to isocurvature perturbation from CMB data are now quite stringent, and constrain the ratio $f/H_{I}$, with implications for the amplitude of inflationary gravitational waves. The complementarity between isocurvature and tensor modes in axion models is explored more fully (for the harmonic limit of the $n=1$ potential) in Ref.~\citep{Hlozek:2017zzf}. In future work, we plan to explore the phenomenology of and constraints to ULA isocurvature perturbations for the much more general class of models considered here.

\subsection{Time evolution of ULA perturbations}

For a fixed wavenumber $k$ there are three time-scales that are important for the ULA mode evolution: i) horizon crossing $k = a_k H(a_k)$; ii) the redshift $z_c$ at which the field starts to oscillate around its minimum; iii) the time $a_s$ after which the sound speed is equal to the oscillation-averaged ULA equation of state, $k= a_s \varpi(a_s)$. Note that, $a_c \equiv 1/(1+z_c)$ is always smaller than $a_s$, or in other words the field starts oscillating before its sound speed starts to evolve. This is because, for the field to acquire $c_s^2 < 1$, it must be oscillating. However, for a given $k$, one can potentially have an arbitrary hierarchy between $a_k$ and $a_c$, and $a_k$ and $a_s$.

We wish to explore the mode evolution of different Fourier modes for the three forms of the ULA potential at fixed $z_c$, which we set to be $10^{-4}$. We choose a fraction of the total energy density at $z_c$ in the ULA to be $f_{z_c}= 0.01$. In doing so, the ULA never makes a significant contribution to the total energy density of the universe.  Since the ULA is always sub-dominant, comparing the evolution of the same wave-number leads to equal $a_k$ for each ULA potential. 

For each value of $n$, we can use Eq.~(\ref{eq:xc}) to translate our condition on $z_c$ to a relationship between $\alpha$ and $\Theta_i$. Similarly by specifying $f_{z_c}$ we fix the relationship between $\mu$ and $\Theta_i$.  We are left with one degree of freedom to fully specify the model: the value of the frequency $\varpi_0$, which enters the effective sound speed after the field starts oscillating and is specified by further fixing $\Theta_i$. In the following discussion we arbitrarily set $\Theta_i=\pi/2$ (choosing another value would not affect our conclusions). The resulting theory and parameter values for the specific model discussed in this Section are shown in Table \ref{tab:params}.
\begin{table}[!tbh]
\begin{tabular}{l|c|c|c}
\hline
\hline
Parameter & $n=1$ & $n=2$ & $n=3$  \\
\hline
$\mu$ & $1.54 \times 10^6$ & $1.09 \times 10^6$ & $8.92\times10^{5}$  \\
\hline
$\alpha$ & 0.124 & 0.175 &0.215  \\
\hline
$\varpi_0\ ({\rm Mpc}^{-1}) $& 341.7 & 0.0185 & $1.11 \times 10^{-4}$\\
\hline
\hline
\end{tabular}
\caption{ 
Theory parameters (determined using the translation equations in Sec.~\ref{sec:translation}) for $z_c = 10^{4}$, $f_{z_c} = 0.01$, and $\Theta_i = \pi/2$.} 
\label{tab:params}
\end{table}

\begin{center}
\begin{figure}
    \includegraphics[scale=0.66]{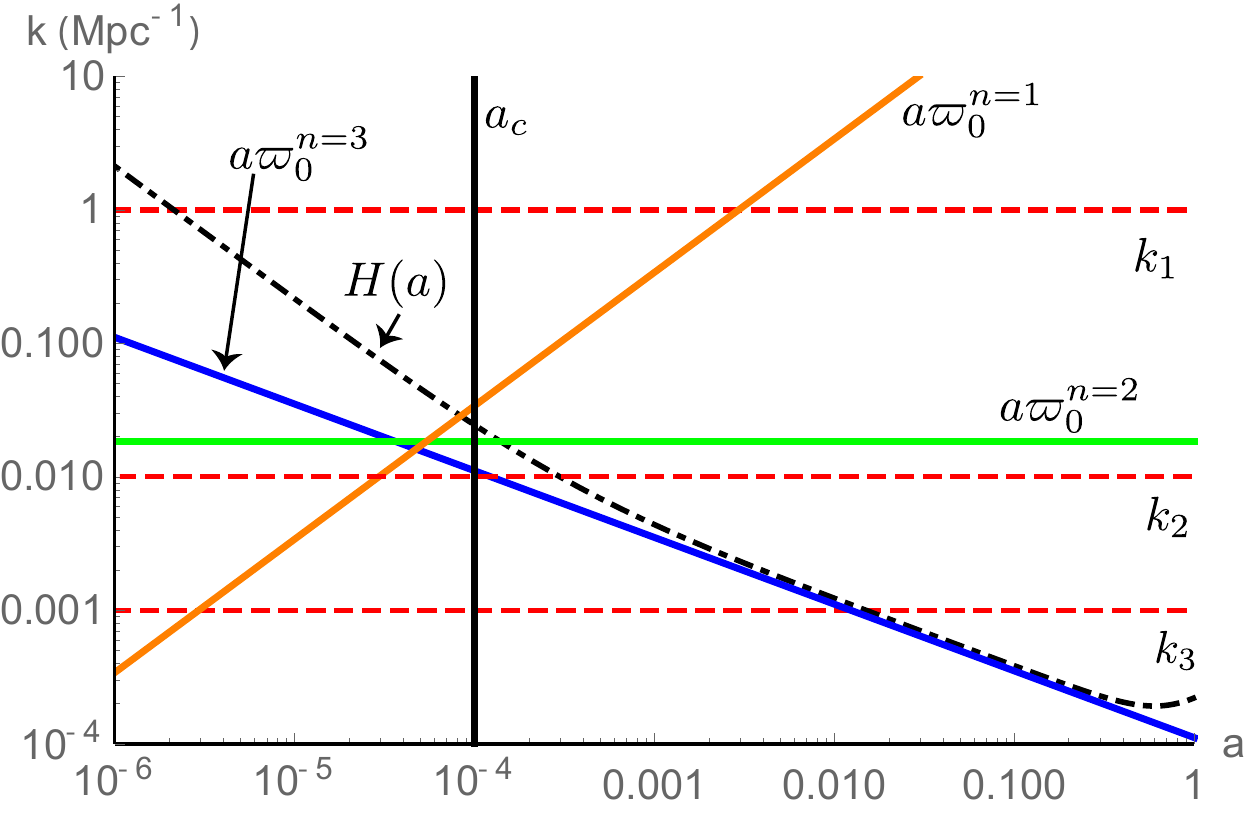}
    \caption{The evolution of a series of scales associated with ULA perturbations. Note that while $k>a \varpi$ the mode has $c_s^2 \simeq 1$.}
    \label{fig:kfig}
\end{figure}
\end{center}

\begin{center}
\begin{figure}
    \includegraphics[scale=0.6]{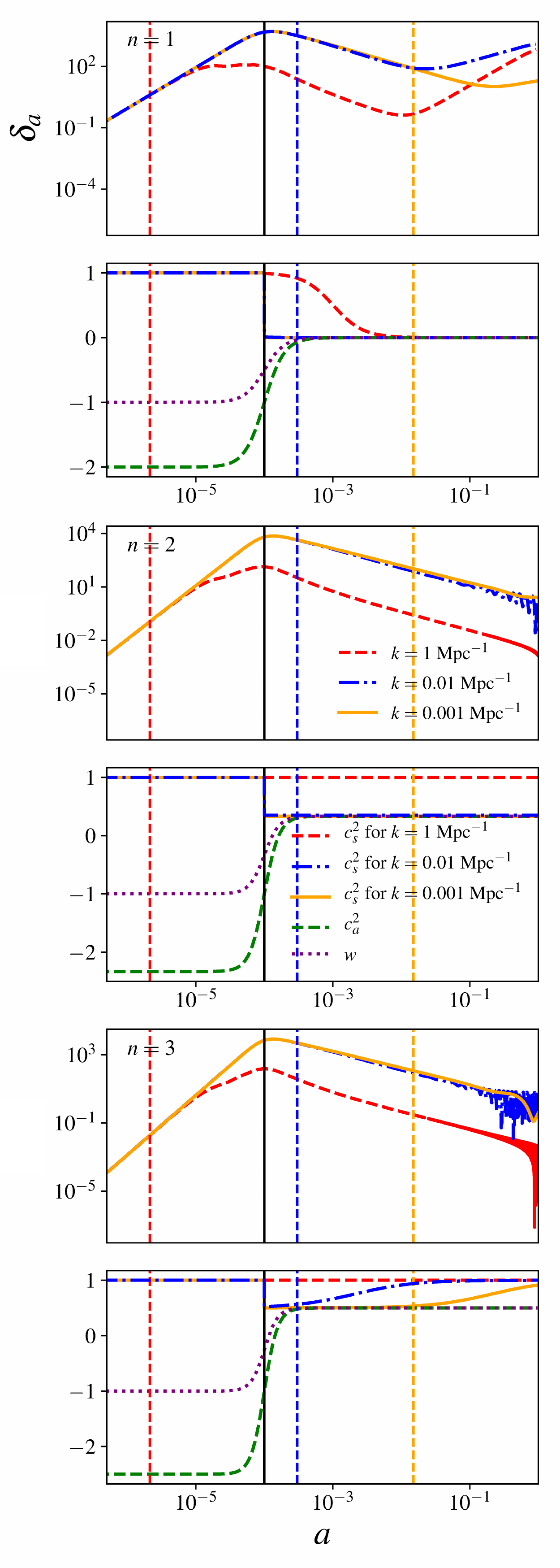}
    \caption{Evolution of the density contrast for the three forms of the ULA potential considered in this paper and with $z_c=10^{4}$, $f_{z_c}=0.01$, and $\Theta_i = \pi/2$. The vertical dashed black line shows $a_c$, while the vertical dashed colored lines show Horizon crossing for each mode.}
    \label{fig:perturb}
\end{figure}
\end{center}

We explore the evolution of three modes: $k_1=1\ {\rm Mpc}^{-1}$, $k_2=10^{-2}\ {\rm Mpc}^{-1}$, and $k_3 = 10^{-3}\ {\rm Mpc}^{-1}$. We show these modes, along with other important scales, in Fig.~\ref{fig:kfig}. From Eq.~(\ref{eq:ceff2}) we can see that if $k\gg a\varpi$ the sound speed goes to $1$. Hence, for a fixed $k$, the effective sound speed evolves differently along time depending on the value of $n$: $c_s^2$ goes from one to less than one for $n=1$, it is a constant for $n=2$, and it evolves from less than one to one for $n=3$.  These modes were chosen because they have different hierarchies: $k_1$ has $a_{k_1}<a_c<a_{s_1}$, $k_2$ has $a_c<a_{k_2}$ and no $a_{s_2}$, and $k_3$ has $a_c<a_{s_3}<a_{k_3}$. In Fig.~\ref{fig:perturb} we show the evolution of the ULA density contrast for these three modes.

At early times, as long as the mode is superhorizon and $a<a_c$, we have $w_a \simeq -1 + c_n (a/a_c)^{3(1+w_n)}$, where $c_n$ is a factor of order unity. The evolution of density perturbations is similar for each ULA potential and each mode, as dictated by the initial behavior in Eqs.~(\ref{eq:deltaSH}) and (\ref{eq:thetaSH}). Since both the density contrast and the heat flux are proportional to $1+w_a$, this shows that for a fixed $a_c$ we expect that the lower values of $n$ will have larger perturbations. This is indeed the case in Fig.~\ref{fig:perturb}.
 
 As illustrated by the wavenumber $k_1$, modes with $a_k < a_c$ enter the horizon while the field is still undergoing slow-roll, EDE, evolution. This results in a suppression in the growth of the perturbations compared to their superhorizon evolution. Once $a \gtrsim a_c$, the field starts to oscillate in its potential and  $w_a \to (n-1)/(n+1)$. As long as $a<a_s$, $c_s^2 = 1$ and the pressure support leads to a strong decrease in the perturbation amplitude. This suppression is present for both superhorizon and subhorizon modes.

Once $a > a_s$, $c_s^2 \to w_a$ and the field's internal pressure support will decrease. In the case where $n=1$ the field is effectively pressure-free and the density perturbation starts tracking that of CDM. For $n>1$ some residual pressure support remains, leading to rapid oscillations in the ULA's density perturbations with an oscillation frequency and amplitude that differs for each $n$ and $k$. 

\section{Impact of an ultra-light axion on the CMB and matter power spectra}
\label{sec:observables}

\begin{figure*}[h]
    \centering
     \includegraphics[scale=0.22]{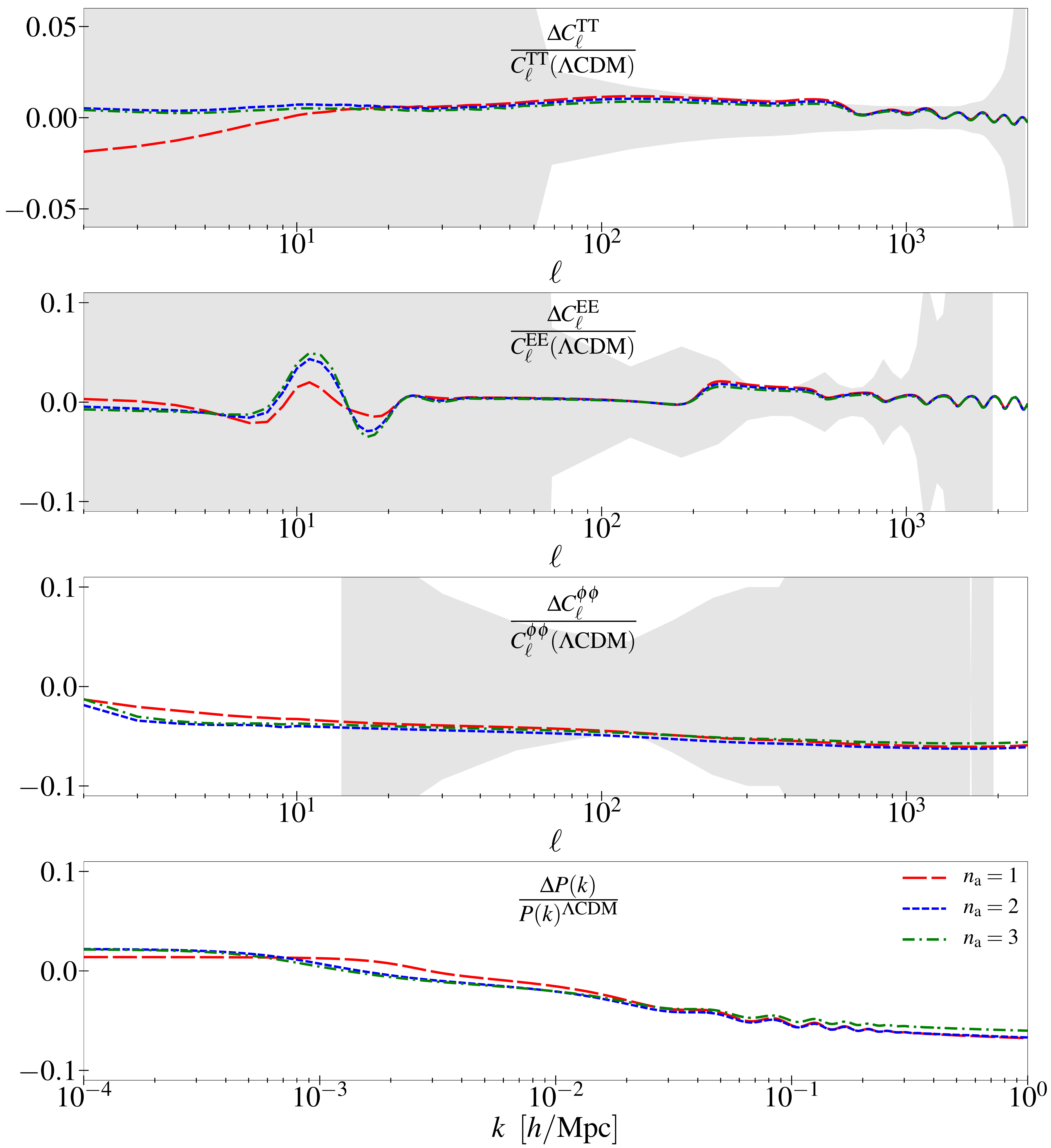}
    
    \caption{Same legend as Fig.~\ref{fig:Cl_residuals_zc10} for $1+z_c = 10$. }
    \label{fig:Cl_residuals_zc10}
\end{figure*}

We compute the CMB and matter power spectra using \texttt{CLASS} for several values of the potential exponent $n=(1,2,3)$ and decay redshift $1+z_c = (10,10^5)$. We set the six $\Lambda$CDM parameters to their best fit values of {\em Planck} TT,TE,EE+lowP 2015 \citep{Ade:2015xua}. We fix the angular scale of the sound horizon, $\theta_s$, which requires us to adjust the value of $H_0$ (this is done using a shooting method implemented in \texttt{CLASS}). We set the density of ULAs to its upper limit at 95\%~C.L. derived in the next Section. The results are shown in  Figs.~\ref{fig:Cl_residuals_zc1e5}  and \ref{fig:Cl_residuals_zc10}.

\subsection{The CMB power spectra in the presence of a ULA}
\label{eq:cps}

\begin{figure*}[h]
    \centering
   \includegraphics[scale=0.22]{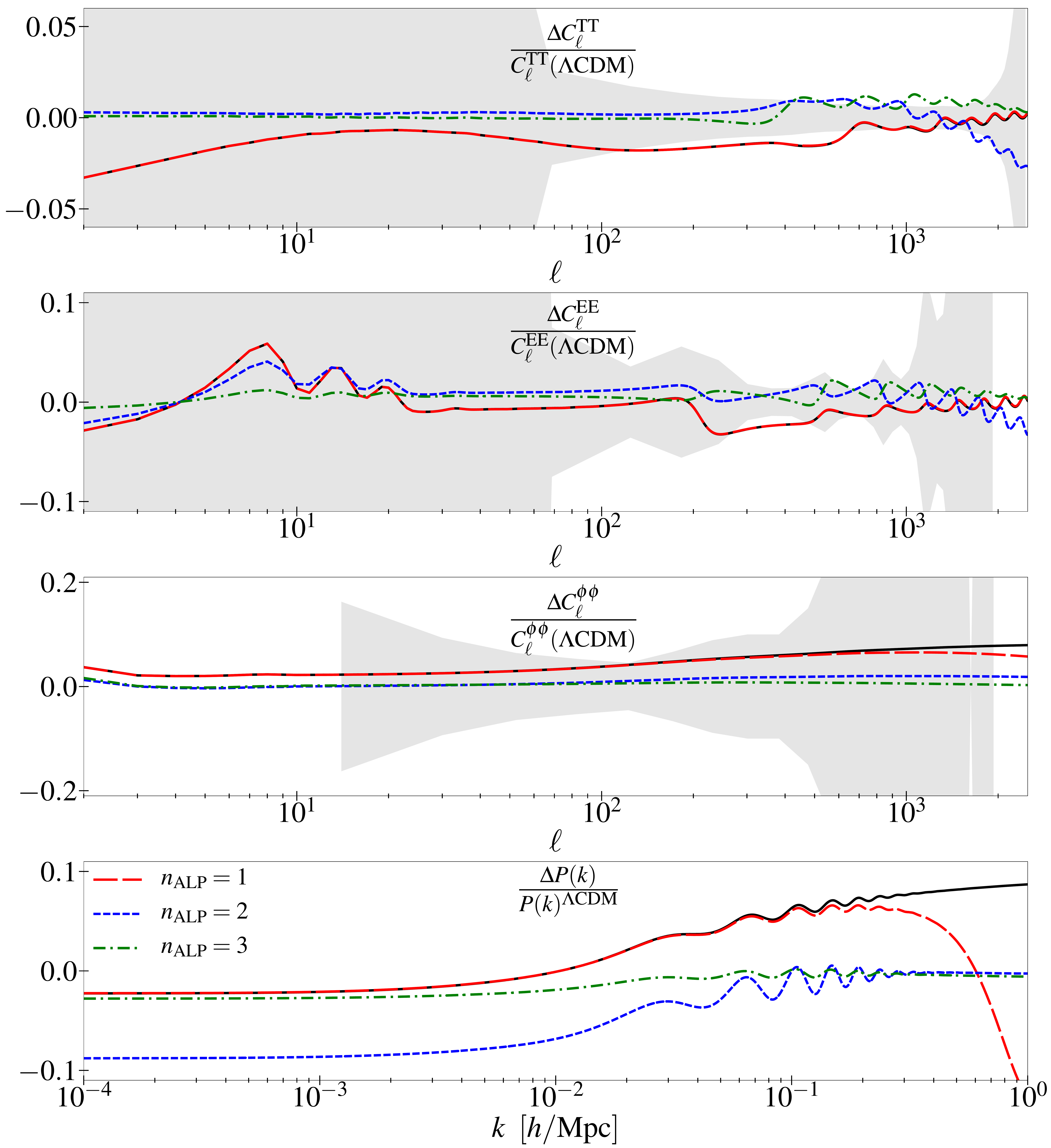}
    \caption{Residuals of the (lensed) CMB TT power spectrum (first panel),  EE power spectrum  (second panel), lensing power spectrum (third) and matter power spectrum (fourth panel) computed for several values of the potential exponent $n=(1,2,3)$ and $1+z_c = 10^5$. Residuals are taken with respect to the $\Lambda$CDM model, with parameters given by the best fit of {\em Planck} TT,TE,EE+lowP \citep{Ade:2015xua}. Axion densities are set at their constraints at 95\% C.L.. The grey bands show {\em Planck} 1$\sigma$ sensitivity.}
    \label{fig:Cl_residuals_zc1e5}
\end{figure*}

If the dilution starts much before matter-radiation equality (Fig.~\ref{fig:Cl_residuals_zc1e5}), CMB power spectra show clear differences depending on the value of $n$, i.e., on the properties of the ULA once it starts diluting.

For $n = 1$, the effects of the ULA are very similar to that of an extra matter component. We illustrate this by comparing it to a universe with an additional pure CDM component, represented by the black curve on each panel of Fig.~\ref{fig:Cl_residuals_zc1e5}. At the highest multipoles, the amplitude of the acoustic peaks is altered by an earlier epoch of matter/radiation equality, changes to the gravitational driving of acoustic oscillations (affecting the Sachs-Wolfe term), and modifications to the blue shift of photons in decaying gravitational wells, the early integrated Sachs-Wolfe (EISW) effect. 
Since we hold the angular sound horizon $\theta_s$ fixed the angular scale $\theta_d$ of diffusion damping will vary and this leads to an altered damping tail. Keeping $\theta_s$ fixed for a flat universe required adjusting the value of $\Omega_\Lambda$, changes the Late Integrated Sachs-Wolfe (LISW) plateau visible at low-$l$'s. It is also visible in the EE spectrum as the reionization history is affected by a change in $\Omega_\Lambda$. Note that the effect of this ULA on the CMB power spectra makes it a viable CDM candidate (see, e.g., Ref.~\citep{Marsh:2010wq}): had we adjusted $\omega_{\rm cdm}$ accordingly, the remaining effects would have been due to a suppression of the matter power spectrum on small scales (which we comment on later) and therefore almost invisible in the CMB, aside from a moderately altered lensing power spectrum.

For $n=2$, the effects of the ULA are very similar to that of an extra radiation component \citep{Hou:2011ec}.  CMB anisotropies are then altered for two reasons. First, at the background level, the additional relativistic species shift matter-radiation equality, which produces modified gravitationally driven oscillations in the photon-baryon plasma and EISW. Hence, the main background effect is due to the requirement that $\theta_s$ is kept fixed and manifests as a shift in the damping scale $\theta_d$, a very mild LISW effect and some oscillation patterns due to different reionization history in the EE spectrum. Second, at the level of perturbations, such an ULA produces a BAO phase-shift distinct from that of true free-streaming particles like neutrinos \citep{Baumann:2015rya}. {\em Planck} data are not only sensitive to the background effect of neutrinos, but also to the ``neutrino-drag" \citep{Smith:2011es,Sendra:2012wh,Audren:2014lsa,Follin:2015hya,Sellentin:2014gaa}, and have already been used to constrain the effective sound speed $c_s^2$ and viscosity $c_{\rm vis}^2$ of the non-CMB radiation component and found to be consistent with that of free-streaming neutrinos \citep{Hu:1998kj}. These parameters are distinct in ULA models, and so we do not expect strong degeneracies between ULAs and neutrinos. 

For $n=3$, the energy density of the axion dilutes faster than any known cosmological species. This leaves less of an imprint on the CMB than the $n=1$ or $n=2$ cases, and most of the effects can be attributed to the EDE phase, rather than to the diluting fluid which becomes quickly invisible.
Since $\theta_s$ is kept fixed, the most important effect of the extra amount of expansion is to reduce the amplitude of the damping tail. On the other hand, the non-adiabatic sound speed of the diluting ULA also leads to small peculiar phase-shift of the acoustic peaks, in a manner different from that of $n=2$ or a free-streaming species. 

If the dilution begins after recombination, the exponent $n$ has much less impact. The EDE phase has a slight impact on the growth of metric potentials around recombination, which leads to features at high multipoles (and especially around $\ell \sim 300$). The additional residual wiggles at high-$l$'s are mostly due to the different amount of lensing. It depends on the impact of the ULA on the matter power spectrum which we comment on below. The difference between the dynamics of the perturbations are mostly visible at small $l$'s. Since we keep $\theta_s$ fixed, $\Omega_\Lambda$ is changed which in turn affects $z_\Lambda$. However, in the $n=1$ case the additional matter component shifts $z_\Lambda$ further, in turn affecting more strongly the LISW plateau, in a manner similar to massive neutrinos. Further differences can be attributed to the impact of the different $w(a)$ as the fluids dilute differently, but fall well below cosmic variance. However, we expect that experiments sensitive to late-time expansion (e.g. JLA, BAO) are sensitive to these effects.

\subsection{The matter power spectrum in the presence of a ULA}
\label{sec:mps}
We now turn to the matter power spectrum, which also shows interesting features strongly dependent on the EDE dilution time and potential power-law index $n$. In general, once $A_s$ and $n_s$ are fixed, the matter power spectrum depends on:  i) the sound horizon at baryon drag $r_s(z_{\rm drag})$ which dictates the phase of the BAO ; ii) the Hubble scale at matter radiation equality $k_{\rm eq}\equiv a_{\rm eq}H_{\rm eq}$ which sets the position of the peak; iii) the ratio $\omega_b/\omega_{\rm cdm}$, which affects the power on scales $k> k_{\rm eq}$  and the contrast of the BAO; iv) the ratio $[g(a_0, \Omega_m)/\Omega_m]^2$ which dictates $k< k_{\rm eq}$ and where $g(a, \Omega_m)=D(a)/a$ is a function expressing how much the growth rate of structures $D(a)$ is suppressed during $\Lambda$
domination. 

When the dilution starts before matter-radiation equality, the ULA affects  $z_{\rm eq}$ especially if it dilutes like matter or radiation. The peak position $k_{\rm eq}$ therefore depends on $n$. If $n=1$, the ratio $\omega_b/\omega_{\rm cdm}$ decreases which leads to a large increase for $k> k_{\rm eq}$  until the mode-dependent sound speed of the ULA kicks in. This creates a turnover at $k> k_{\rm eq}$ that is very specific to such a ULA. One can see that the only difference between a pure CDM component and a ULA is this cutoff on very small sales because of the non-zero pressure support. For $n=2$ and $n=3$, this branch is almost unaffected for such small values of $\Omega_a$. The small-$k$ branch on the other hand is affected by the increase in $\Omega_M = 1-\Omega_\Lambda$ (decrease in $\Omega_\Lambda$) that is required to keep $\theta_s$ fixed. Moreover, for all values of $n$ the BAO is shifted because of different $r_s(z_{\rm drag})$.
  
When the dilution starts after matter-radiation equality, the effects are very similar to that of massive neutrinos, and manifests in two ways. First,  the ratio $k_{\rm eq} /(a_0 H_0)$ governing the location of the maximum in the matter power spectrum depends on the duration of matter domination. Any modification of this ratio leads to an overall shift of the spectrum. It is affected by the presence of an EDE, but the additional matter component (for the $n=1$ case) partially counteracts the effect of the EDE. Hence, the power spectrum in the $n=2$ and 3 case is shifted in the same way, and slightly more than in the $n=1$ case. Second, the additional pressure support leads to suppression of power on small scales in a manner that depends on each fluid sound speed, and thus differs for each $n$. 

\section{Current constraints to ULAs}
\label{sec:constraints}

Using current measurements of the CMB and other probes of large-scale structure we place constraints on the energy density of ULAs as a function of the time when they become dynamical. As mentioned before, although the CMB decouples around $z\sim 1000$, each multipole carries with it information about the evolution of the universe around the time the scales that form it entered the causal horizon. This, in principle, makes the CMB sensitive to cosmological dynamics as long ago as $z \sim 10^5-10^6$ \citep{Linder:2010wp, Karwal:2016vyq}. 

To perform this analysis we consider a series of fixed values for $z_c$ at which we constrain the energy density in the ULA. In addition to this we assume a uniform prior on the initial field value, $\Theta_i$, which in turn implies a particular prior on the ULA's oscillation frequency today, $\varpi_0$ [see Eq.~(\ref{eq:omegan})]. 

\subsection{Description of the data sets and analysis}

We run Monte Carlo Markov chains using the public code {\sc Monte Python} \citep{Audren:2012wb}. 
We perform the analysis with a Metropolis Hasting algorithm, assuming flat priors on $\{\omega_b,\theta_s,A_s,n_s,\tau_{\rm reio},\omega_{\rm cdm}\}$ and a logarithmic prior on $\Omega_{\rm a}$. We scan over 9 points in $1+z_c$ logarithmically distributed between 1 and $10^8$. We also vary $n$ to be equal to $(1,2,3)$.
We make use of {\em Planck} high-$l$ and low-$l$ TT,TE,EE and lensing likelihood. We include the anisotropic BAO data at $z=0.2-0.75$ from the BOSS DR12 data release \citep{Alam:2016hwk} and isotropic BAO data at $z= 0.105$ \citep{Beutler:2011hx} and $z=0.15$ \citep{Ross:2014qpa}. We include the Joint Likelihood Analysis (JLA) of supernovae, which includes measurements of the luminosity distance of SN1a up to redshift $z\sim1$ \citep{Betoule:2014frx}.

\begin{figure*}
    \centering
    \includegraphics[scale=0.5]{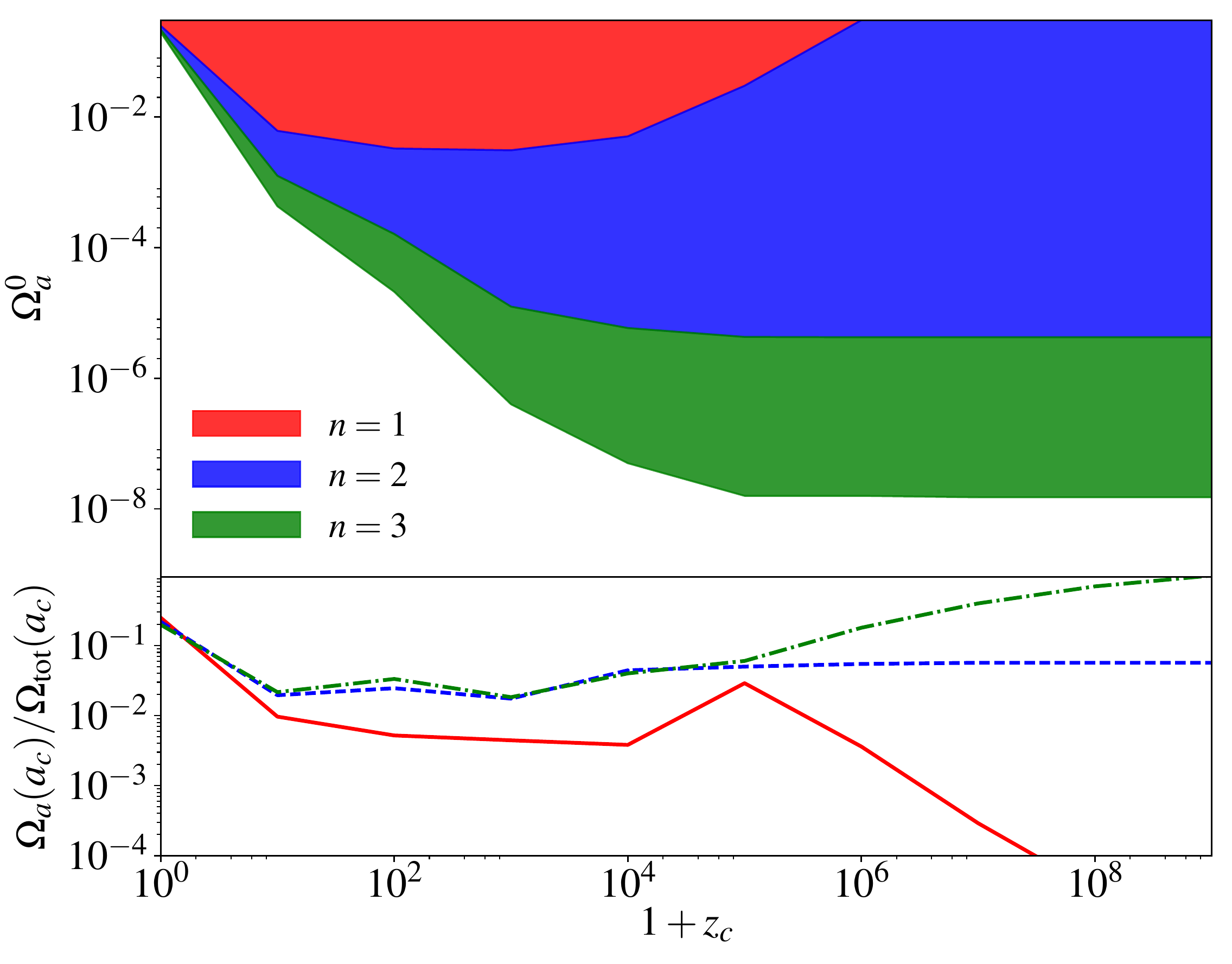}
    \caption{ {\em Top panel $-$} Constraints on the density of the ULA today as a function of its dilution time $1+z_c$. {\em Bottom panel $-$} Constraints on the fraction of the total energy content in the form of a ULA at $a_c\equiv(1+z_c)^{-1}$.}. 
    \label{fig:constraints_today}
\end{figure*}

Although not specified here for brevity, there are many nuisance parameters that we analyze together with the cosmological ones. To this end, we make use of a Choleski decomposition which helps in handling the large number of nuisance parameters \citep{Lewis:2013hha}. We consider chains to be converged using the Gelman-Rubin \citep{Gelman:1992zz} criterion $R -1<0.05$. 
The constraints on the density of ULAs today as a function of their dilution redshift $1+z_c$ are shown in Fig.~\ref{fig:constraints_today}. These have the characteristic `belly' or U-shape first estimated in Refs. \citep{Frieman:1995pm,Amendola:2005ad}, then generated more robustly from a Boltzmann code with MCMC methods in Ref.~\citep{Hlozek:2014lca}, and confirmed in Ref.~\citep{Hlozek:2017zzf} \footnote{This shape seems to be somewhat generic in models for which a species behaves as something other than matter up until a critical transition redshift $z_{c}$. For example, if the dark matter is generated at late times by the decay of a relativistic species, as in the late-forming dark matter model of Ref.~\citep{Sarkar:2014bca}, a \textit{qualitatively} similar constraint plot results.}.

\subsection{Late time constraints}
\label{sec:late_time_constraints}

Constraints on the ULA at late times are driven by measurements of the luminosity distance up to $z\simeq 1$ using the JLA data set \citep{Betoule:2014frx} and angular diameter distance \citep{Alam:2016hwk,Beutler:2011hx,Ross:2014qpa}. Note that even for $z_c=0$ the field evolves away from $w_\phi =-1$- in particular, fitting the parameterization $w_\phi(z) = w_{a,0} + w_{a,1}[1-1/(1+z)]$ to the three forms of the potential gives \begin{eqnarray}
n=1 \rightarrow w_{a,0} = -0.50,\ w_{a,1} = -0.79,\\
n=2 \rightarrow w_{a,0} = -0.37,\  w_{a,1} = -1.18,\\
n=3 \rightarrow w_{a,0} = -0.31,\  w_{a,1} = -1.36.
\end{eqnarray}
The values of these parameters show the behavior we expect as a function of $n$: as $n$ increases the scalar field's energy density decreases more rapidly, leading to a smaller $w_{a,0}$ and $w_{a,1}$ with increasing $n$. 

The JLA data (combined with measurements of the temperature anisotropy from {\em Planck}, polarization measured by WMAP and measurements of the BAO) yield a constraint of $w_0 = -0.957 \pm  0.124$ and $w_1 = -0.336 \pm 0.552$ \citep{Betoule:2014frx}, where $w(z) = w_0 + w_1 [1 - 1/(1 + z)]$. If we choose a small value of $z_c$ then the ULA will behave as quintessence and contribute to driving the current epoch of accelerated expansion. Fixing the matter component at $\Omega_{m,0} = 0.3$ the equation of state of the late-time dark sector (consisting of $\phi$ and a cosmological constant) is given by 
\begin{eqnarray}
    w(z) &=& \frac{-1 + w_a \rho_a/\rho_\Lambda}{1+ \rho_a/\rho_\Lambda},\\
    &=& \frac{-1+w_a(z) \Omega_a(z)/(0.7-\Omega_{a,0})}{1+\Omega_a(z)/(0.7-\Omega_{a,0})} \nonumber
\end{eqnarray}
Note that the cosmological constant plus ULA dark sector has $w\geq -1$.

\begin{figure*}
    \centering
    \includegraphics[scale=0.27]{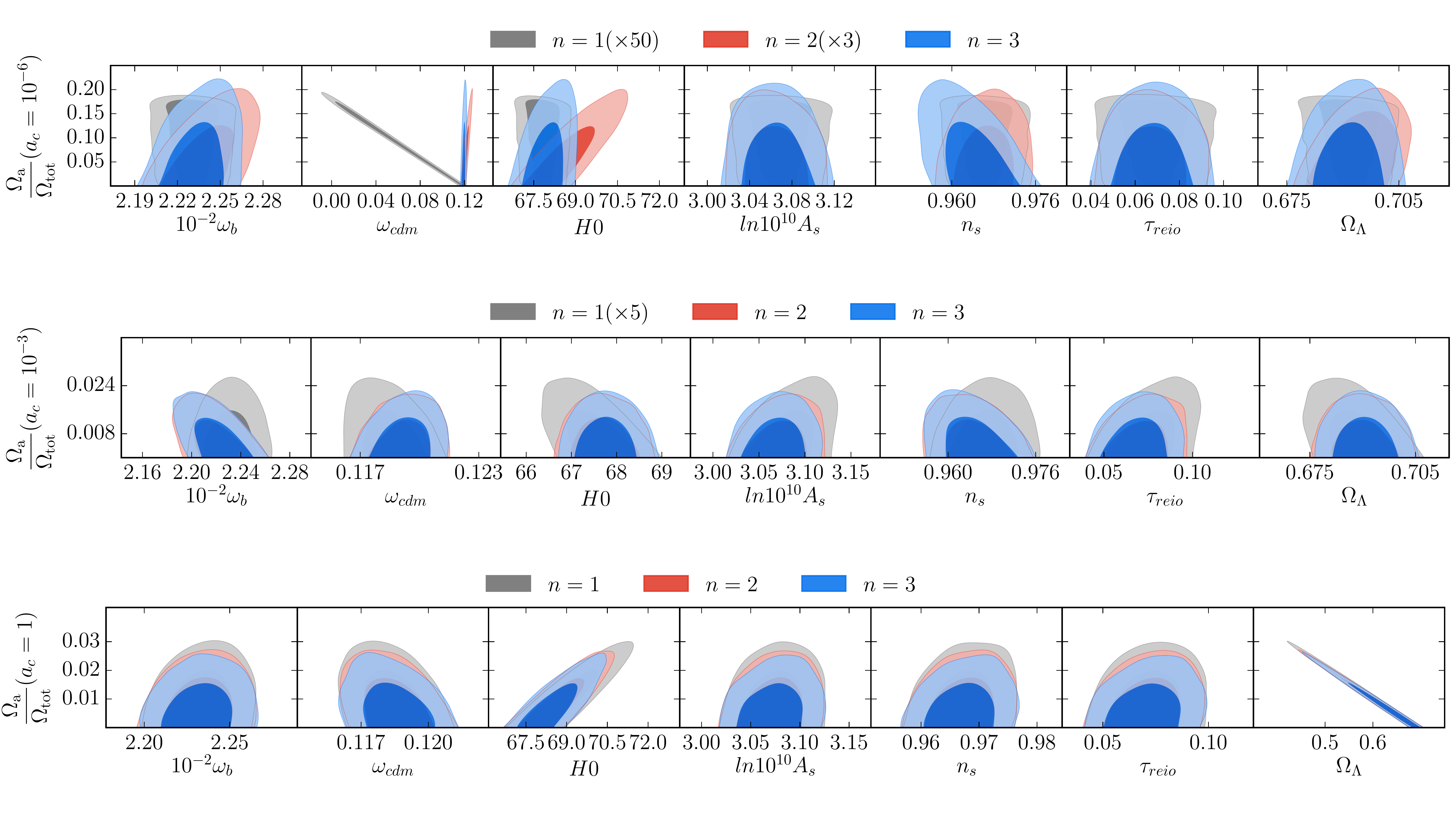}
    \caption{ Posterior distributions of the density of ULAs today vs the other $\Lambda$CDM parameters for $1+z_c = 1$ (bottom panel), $10^3$ (middle panel) and $10^6$ (top panel). } 
    \label{fig:triangle}
\end{figure*}

We can then use this equation of state and fit for $w_0$ and $w_a$ up to $z=1$ to find the JLA-driven constraint on a late-time ULA. For example, if $z_c = 0$ then we find that constraints to ULAs are driven by the fact that $w_0<-0.833$ and we find that at a 68\% CL we have
\begin{eqnarray}
n=1 &\rightarrow& \Omega_{a,0}<0.22,\\
n=2 &\rightarrow& \Omega_{a,0}<0.16,\\
n=3 &\rightarrow& \Omega_{a,0}<0.15.
\end{eqnarray}
This discussion also shows that we expect the JLA constraint to give a degeneracy between $\Omega_{a,0}$ and $\Omega_{\Lambda}$ such that $\Omega_{a,0}+\Omega_{\Lambda} = 1-\Omega_m \simeq 0.7$.  
This simple estimate is very close, albeit slightly stronger, than what is obtained in a real analysis:
\begin{eqnarray}
n=1 &\rightarrow& \Omega_{a,0}<0.25,\\
n=2 &\rightarrow& \Omega_{a,0}<0.22,\\
n=3 &\rightarrow& \Omega_{a,0}<0.20.
\end{eqnarray}

Note that there are some additional effects on the CMB (LISW, reduced lensing amplitude) which are well below {\em Planck} sensitivity but could be probed by future experiments.

\subsection{Constraints for $z_c$ around recombination}
\label{sec:z_c_recomb_constraints}

When the dilution begins after matter-radiation equality but before recombination, the $n=2$ and $n=3$ cases are basically identical; indeed, the very fast diluting fluid leaves no significant additional impact on the CMB as the universe is largely matter dominated by then. Hence, the constraints are purely driven by the EDE phase. The strongest degeneracy visible on Fig.~\ref{fig:triangle} -- middle panel --  appears to be with $\omega_b$, which can be adjusted to counteract the effect of a faster recombination. Additional mild degeneracies appear with parameters governing the overall shape of the power spectrum \{$A_s\exp(-\tau_{\rm reio})$,$n_s$\} and the amplitude of the EISW term ($\omega_{\rm cdm})$. Note that $H_0$ shows no degeneracy with $f_a(z_c)$. In fact $z_c =10^3$ represents a turning point in the direction of the degeneracy; for higher value of $z_c$, the correlation is {\em positive}, and can be understood in the same manner as the degeneracy between an additional ultra-relativistic species and $H_0$ (e.g.~Ref.~\citep{Hou:2011ec}). For lower values of $z_c$ however, the correlation becomes {\em negative} and is driven by the requirement of keeping the angular size of sound horizon at recombination $\theta_s$ fixed.

The $n=1$ case, however, represents a very distinct case: as the fluid dilutes like matter, it increases the total matter component of the universe. Hence, the constraints are driven by the additional matter component and degeneracies with $\Lambda$CDM parameters can be understood accordingly. As expected, a strong negative degeneracy appears with $\omega_{\rm cdm}$, as well as with $\Omega_\Lambda = 1-\Omega_M$ (valid in a flat universe) because $\Omega_M$ increases. Similarly to the previous case, some mild degeneracies appear with \{$A_s\exp(-\tau_{\rm reio})$,$n_s$\} as to compensate the overall shape of the spectra. Finally, a strong negative correlation appears with $H_0$ and is due to purely geometric effects: one needs to compensate the increase in the Hubble rate ($\propto \sqrt{\Omega_M(1+z)^3}$) by decreasing $H_0$ in order to keep the same angular diameter distance to recombination \citep{Hlozek:2014lca}.

\subsection{Constraints for $z_c$ earlier than matter-radiation equality}
\label{sec:before_z_eq_constraints}

We have described in Sec.~\ref{sec:observables} the effect of an early dilution on the CMB power spectra, well before matter radiation equality. The degeneracies visible on Fig.~\ref{fig:triangle} -- bottom panel --  are straightforward to understand. First and foremost, when $z_c\gtrsim10^5$ and $n=1$, the ULA becomes fully degenerate with a matter component. This represents a range of mass for which the axion is a valid DM candidate, as pointed out in Refs.~\citep{Hlozek:2014lca,Marsh:2015xka}. Note that the degeneracy is not perfect at $z_c = z_{\rm eq}$; this is because this requirement does not ensure that $z_{\rm eq}$ is exactly fixed, $z_c$ represents a transition redshift and the fluid does not behave exactly like matter at that time. Moreover, the CMB is sensitive to details of the expansion history around matter-radiation equality through the EISW, which further increases the value of $z_c$ at which the ULA is degenerate (in CMB observations) with the CDM component.
 Note that there are no strong degeneracies between the $n=1$ ULA and any other cosmological parameters in this case: this is expected because $\omega_{\rm cdm}$ shows no strong degeneracy with any parameters within $\Lambda$CDM.
 
In the $n=2$ case, the fluid dilutes like an extra radiation component: the constraint is therefore driven by this additional relativistic species. As explained in Sec.~\ref{sec:observables},  we expect a degeneracy with $N_{\rm eff}$ to some extent. Indeed, at the background level if the dilution starts early enough ($z\gtrsim10^6$), they have exactly the same behaviour.  However, we confirm that the degeneracy is far from perfect because perturbations in the ULA fluid are very different from that of a free-streaming species like neutrinos. The ULA has a scale-dependent sound speed and viscosity that differs strongly from that measured by {\em Planck} high-$\ell$ TT,TE,EE+low-P \{$c_s^2 = 0.3240 \pm 0.0060,c_{\rm vis}^2 = 0.327 \pm 0.037$\} \citep{Ade:2015xua} and therefore cannot replace the totality of the non-CMB radiation bath. We find that it can account at most for $\sim20\%$ of the total $N_{\rm eff}$.
Degeneracies with other parameters can be understood in a similar way as that of an additional relativistic species (e.g., Ref.~\citep{Hou:2011ec}), and we comment in Sec.~\ref{sec:implications_for_tensions} on the strong correlation with $H_0$.

Finally, for $n=3$ the constraints come mostly from the EDE phase and are thus very similar to that of $z_c\sim 10^3$. In particular it is straightforward to show that if CMB measurements constrain the fractional ULA contribution at $z_{\rm CMB}$ to be less than $f_{\rm CMB}$ then as $z_c \gg z_{\rm CMB}$ the limit on $\Omega_{a,0}$ asymptotes to 
\begin{equation}
    \Omega_{a,0} < \frac{f_{\rm CMB} \Omega_{r,0}}{(1-f_{\rm CMB})\sqrt{(1+z_{\rm CMB})}}\,.
\end{equation}
Taking $\Omega_{r,0} = 9.2 \times 10^{-5}$ for photons, three massless neutrinos and $h=0.68$ and setting $z_{\rm CMB} = 10^5$ and $f_{\rm CMB} = 0.06$ we find that the asymptotic constraint to $\Omega_{a,0}$ for a ULA with $n=3$ is approximately $ \Omega_{a,0} <2 \times 10^{-8}$ which agrees well with the constraints shown in Fig.~\ref{fig:constraints_today}. We can also translate this into a constraint on the $f_{z_c}$:
\begin{equation}
    f_{z_c} < \frac{f_{\rm CMB}\sqrt{\frac{1+z_c}{1+z_{\rm CMB}}}}{1+f_{\rm CMB}\left(\sqrt{\frac{1+z_c}{1+z_{\rm CMB}}}-1\right)}\,.
\end{equation}
This expression shows that with the current constraint $f_{\rm CMB}=0.06$ at $z_{\rm CMB}=10^5$ we limit $f_{z_c}$ to be less than unity as far back as $z_c = 10^{10}$. At this time constraints on the rate of expansion of the universe during big bang nucleosynthesis from measurements of the primordial light element abundances can, in principle, be used to further restrict $f_{z_c}$ (see, e.g., Ref.~\citep{Carroll:2001bv}). 

When looking at the degeneracies between the $n=3$ ULA and other cosmological parameters the only difference relative to the other cases is with $\omega_b$, for which the degeneracy is flipped: in that case, the EDE does not affect the recombination physics, but it decreases the damping tail of the CMB. This effect can be partially compensated by increasing $\omega_b$. This fact also drives the degeneracy with $n_s$. Note that, as $z_c$ increases, the constraint on $\Omega_a$ today flattens: this means that the constraints on $f_a(z_c)$ {\em relaxes} as $z_c$ increases. This is expected because {\em Planck} (limited to $\ell < 2500$) is less and less sensitive to physics above $z\sim 10^5$.

\subsection{The role of perturbations}

Finally, we comment on the extent to which the details of perturbations play a role in the constraining power. In previous discussion, we have seen that the $n=1$ case is purely degenerate with a CDM component if $z_c \gtrsim 10^5$; naturally this degeneracy disappears if we neglect perturbations in the fluid. However, when $z_c\lesssim 10^4$, we find that neglecting perturbations leads to constraints that differ by no more than $\sim$20\%. In the $n=2$ and $n=3$ case, we find a similar difference. 

Note however, that in the $n=2$ case conclusions would have changed if the perturbations of the ULA were that of a free-streaming species. In that case, we would have found a perfect degeneracy for high-enough $z_c$ that would not have been present if  perturbations are neglected. It is only because the ULA is constrained to be a sub-dominant fraction of the universe components (and thus never drives the expansion and evolution of perturbations), that the details of their perturbations don't matter too much. 

However, in the future, next generation CMB experiments and LSS surveys are expected to improve sensitivity on ULA. Hence, any detection will require an accurate description of perturbations, potentially even beyond the fluid approximation described in this paper. In future work, we will investigate the accuracy of this approximation compared to a full solution of the KG equation, with an eye towards the sensitivity levels of future CMB experiments like CMB-S4 \citep{Abazajian:2016yjj}.

\section{Implications for cosmological tensions}
\label{sec:implications_for_tensions}

Although most cosmological observables are individually consistent with a $\Lambda$CDM cosmology, tensions exist between the predictions of various data sets, such as the Hubble tension \citep{Riess:2016jrr,Ade:2015xua}. 
Furthermore, the recent measurement of the sky-averaged 21-cm signal by the Experiment to Detect the Global Epoch of Reionization Signature (EDGES) is inconsistent with predictions of $\Lambda$CDM \citep{Bowman:2018yin}, although theinterpretation of the signal is still being explored \citep{Hills:2018vyr}. In this section, we examine the effect of ULAs on these two tensions. 

\subsection{The Hubble tension}

One of the most prominent and persistent tensions in cosmology is the Hubble tension \citep{Freedman:2017yms,Bernal:2016gxb}.
The current expansion rate of the universe as predicted by the $\Lambda$CDM model when fit to the CMB disagrees with local measurements at greater than $3\sigma$ \citep{Riess:2016jrr}. Planck determines $H_0$ to be $66.93 \pm 0.62$ km s$^{-1}$ Mpc$^{-1}$, while the SH0ES (Supernova H0 for the Equation of State  Collaboration) collaboration measures a value of $73.24 \pm 1.74$ km s$^{-1}$ Mpc$^{-1}$ \citep{Riess:2016jrr}. 
Numerous explanations have been proposed and studied in the literature \citep{Riess:2016jrr,Karwal:2016vyq,Aubourg:2014yra,Dvorkin:2014lea,Wyman:2013lza,Leistedt:2014sia,Ade:2015rim,DiValentino:2016hlg,DiValentino:2017rcr,DiValentino:2017iww}.

\begin{figure}[h!]
    \centering
     \includegraphics[scale=0.4]{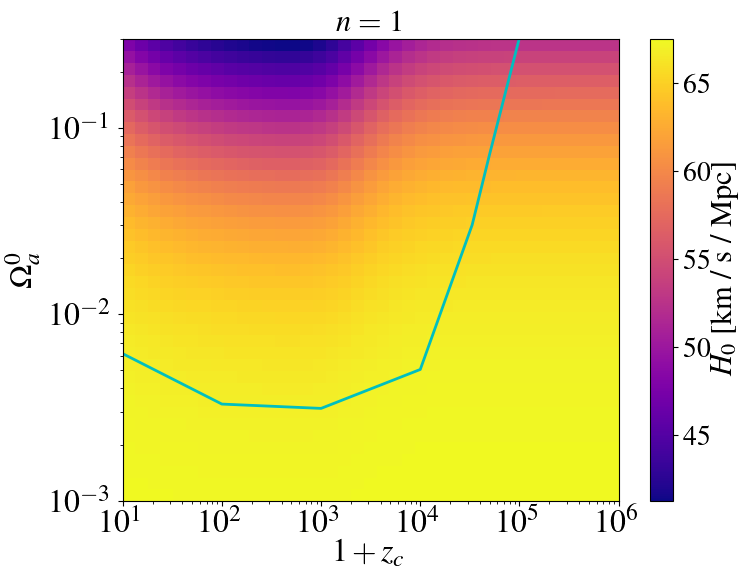}
    \includegraphics[scale=0.4]{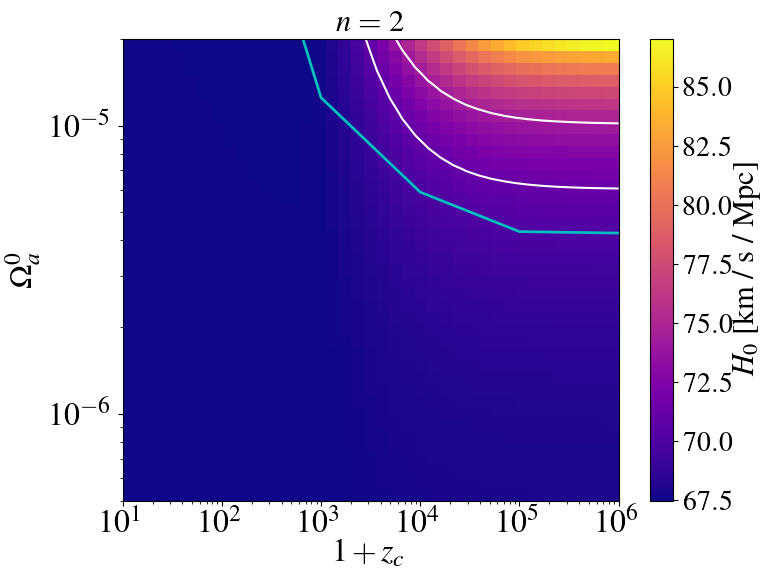}
    \includegraphics[scale=0.4]{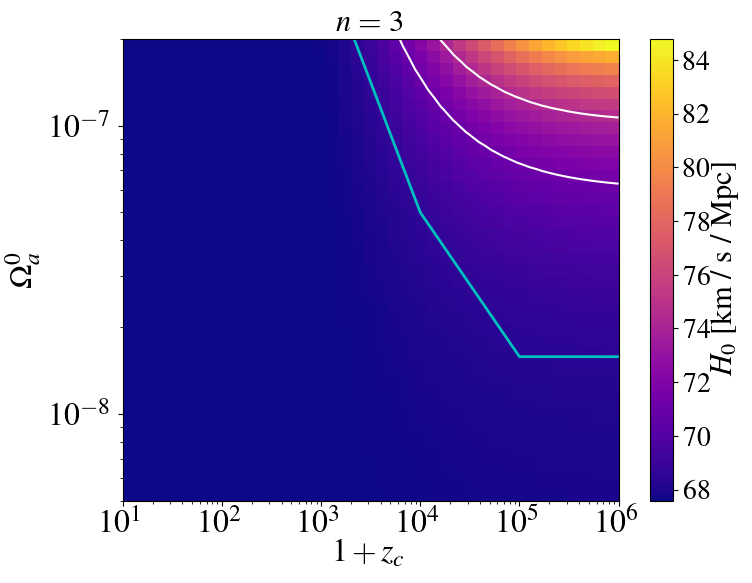}
    \caption{Hubble parameter $H_0$ for various values of $\Omega_a^0$ and $z_c$, for the $n=1$ (top panel), $n=2$ (middle panel) and $n=3$ case (bottom panel). 
    The cyan line represents the constraints shown in Fig.~\ref{fig:constraints_today}.
    The white contours show the $1\sigma$ contour on the $H_0$ value measured by SH0ES.
    } 
    \label{fig:H0_n_2}
\end{figure}

In this section, we investigate whether ULAs can alleviate the tension and what regions in the $\Omega_a - z_c$ plane are best suited to do so, similar to Ref.~\citep{Karwal:2016vyq}. We use the Friedmann equation to compute $H_{0}$ today, given fiducial values for the other cosmological parameters, and the indicated values for $z_{c}$ and $\Omega_{a}$. We keep $\theta_s$ fixed and let \texttt{CLASS}{} solve for the value of $H_0$. The results are shown in Fig.~\ref{fig:H0_n_2}. 

For $n=1$, we find that no value of $\Omega_{a,0}$ for values of $1+z_c \in [10^0, 10^6]$ diminished the $H_0$ tension. With reference to Fig.~\ref{fig:triangle} and Sec.~\ref{sec:before_z_eq_constraints}, for $z_c \gg z_{\rm eq}$, the fluid is fully degenerate with CDM, and $\omega_{\rm cdm}$ and $\Omega_a$ are negatively correlated.
That is, the CMB cannot distinguish between the fluid and CDM. An increase in the energy density of the fluid today will be accompanied by a decrease in the energy density of CDM and there is no change to the value of the Hubble parameter. At the other end, for $z_c \simeq 0$, the fluid is strongly degenerate with $\Lambda$. 
This degeneracy is weaker than that with $\omega_{\rm cdm}$ at $z_c \gg z_{\rm eq}$, because the equation of state parameter of the fluid is not exactly $-1$, as discussed in Sec.~\ref{sec:late_time_constraints}. 
Again, an increase in $\Omega_{\rm a}$ is accompanied with a decrease in $\Omega_{\Lambda}$ and the value of $H_0$ remains unaltered. The tension is, however, somewhat alleviated as the fluid is degenerate with $H_0$ and leads to a larger error on $H_0$. For intermediate redshifts $z_c \lesssim z_{\rm eq}$, the fluid reduces the value of $H_0$, exacerbating the tension. As mentioned in Sec.~\ref{sec:z_c_recomb_constraints}, the angular diameter distance $D_A(z_*)$ to the CMB fixes the value of $\Omega_m h^2$ and therefore, effectively increasing $\Omega_m$ leads to a reduction in $h$. Hence at best, the $n=1$ scenario leaves $H_0$ unaltered, at worst, exacerbates the tension. 

The $n=2$ scenario fares better, as seen from Fig.~\ref{fig:H0_n_2}. For $z_c < z_*$, it fares similarly to the $n=1$ case. It is strongly degenerate with $\Omega_{\Lambda}$ for $z_c \simeq 0$. For $0 < z_c \ll z_*$, the fluid exacerbates the tension. Again, this is due to its effect on $D_A(z_*)$ - it adds to the expansion rate at late times and $H_0$ must decrease to compensate and preserve $D_A(z_*)$. 
For $z_c \simeq 10^3$, as mentioned in Sec.~\ref{sec:z_c_recomb_constraints}, $\Omega_a$ and $H_0$ are uncorrelated. As we are already in matter domination by $z = 10^3$, a fluid that behaves like $\Lambda$ before and radiation after recombination will impact expansion history only over a finite redshift range around $z_c$. 
As the angular diameter distance $D_A(z_*)$ to recombination gets most of its contribution from lower redshifts, its value and therefore $H_0$ remain largely unchanged. For $z_c > z_*$, the $n=2$ scenario is degenerate with $N_{\rm eff}$, as it effectively adds more radiation to the Universe. Hence the impact of the fluid on $H_0$ is similar to that of $N_{\rm eff}$: it increases $H_0$ and diminishes the tension \citep{Riess:2016jrr}. However, our CMB constraints exclude this solution.

Finally, for the $n=3$ scenario, for $z_c \lesssim z_*$, the impact of the fluid is similar to the $n=2$ case.
As mentioned before, the $n=3$ case only impacts expansion history over a small range in redshift centered around $z_c$. For $z_c > z_*$ and $\Omega_a^0$ larger than our current constraints, pre-recombination expansion rate is increased. 
This decreases the radius $r_s$ of the sound horizon at recombination and $H_0$ increases to compensate and preserve $\theta_s$. Hence, the fluid is capable of increasing $H_0$ as seen in Fig.~\ref{fig:H0_n_2}, but for values of $\Omega_a^0$ that are much larger than our constraints. 

The CMB becomes insensitive to physics above $z \sim 10^6$ as noted by \citep{Karwal:2016vyq,Linder:2010wp}. 
Therefore, for a given $\Omega_a^0$, even as $z_c$ increases above $10^6$, the energy density of the fluid for $z \lesssim 10^6$ remains unchanged, as does the Hubble parameter. 
We hence only show the change to the Hubble parameter due to the addition of ULAs up to $1+z_c = 10^6$. 

To summarize, we find that in order for ULAs to diminish the Hubble tension, with $n=2$ and 3, it requires $z_* < z_c \lesssim 10^6$ and $\Omega_a$ larger than our constraints. Although these large values of $\Omega_a$ can solve the tension, they are ruled out  as a fluid with non-adiabatic perturbations shifts the positions of the higher acoustic peaks \citep{Baumann:2015rya}. However we note that it is possible to reduce the tension in the $n=2$ case, making its significance less than 2$\sigma$. We also note that the $n=2$ EDE scenario leads to a more significant easing of this tension than a relativistic species with arbitrary sound speed and viscosity, which can only relax the tension at the $2.4\sigma$ level \citep{Poulin:2018zxs}.

\subsection{EDGES exotic 21 cm measurement}

\begin{figure}[!h]
    \centering
    \includegraphics[scale=0.4]{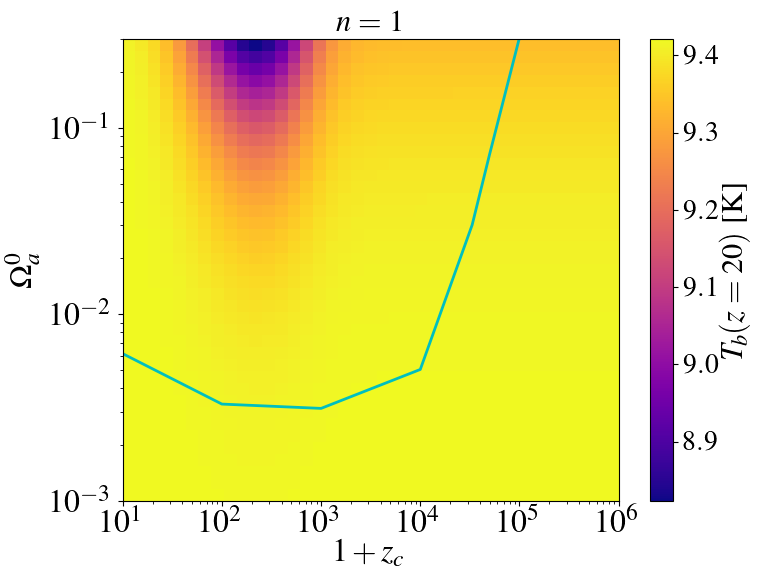}
    \includegraphics[scale=0.4]{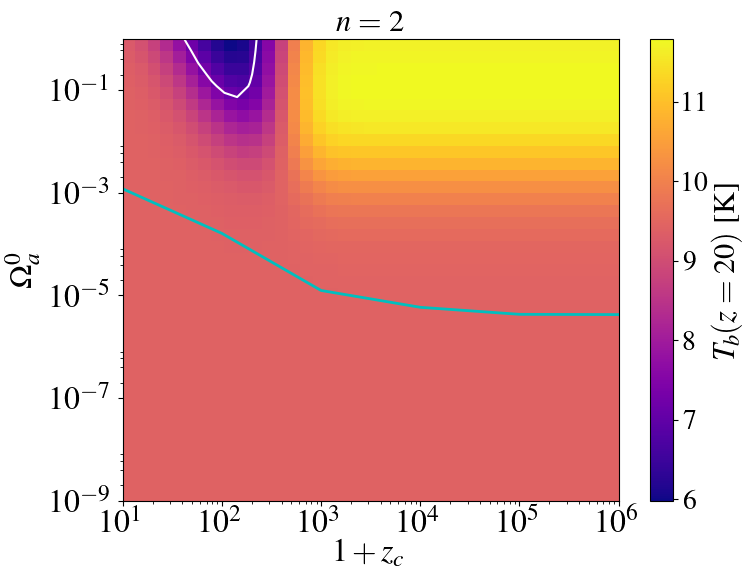}
    \includegraphics[scale=0.4]{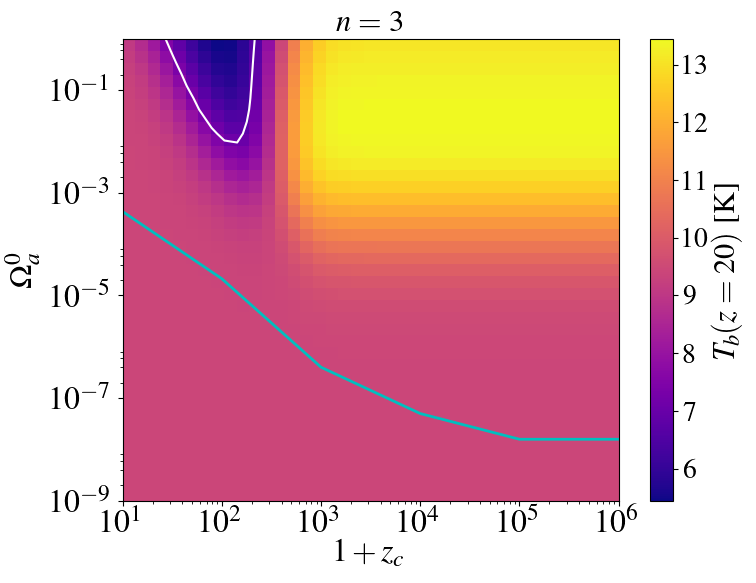}

    \caption{Baryon temperature at $z=20$ (close to the minimum of the absorption trough measured by EDGES \citep{Bowman:2018yin}) as a function of the ULA density today $\Omega_a^0$ and critical redshift $z_c$. The top panel presents the $n=1$ case, the middle panel the $n=2$ case and the bottom panel the $n=3$ case. The white line shows $T_b = 7$ K, i.e. the 99\% upper-limit on the temperature measured by EDGES. The cyan line shows the {\em Planck} 95\% C.L. limit derived in this work. All models of interest are excluded by our analysis.}
    \label{fig:Tb_ula}
\end{figure}

EDGES recently measured the sky-averaged 21cm brightness temperature \citep{Bowman:2018yin} around the redshift range $z=15-20$ to be roughly $2.5$ times smaller ($3.8\sigma$) than that predicted by $\Lambda$CDM. \footnote{The proper interpretation of this this measurement is still under discussion \citep{Hills:2018vyr}.} Two main classes of solution have been suggested to explain this measurement: either the temperature of the photons against which the 21-cm temperature of the gas is measured is brighter than that of the CMB \citep{Ewall-Wice:2018bzf,Mirocha:2018cih,Pospelov:2018kdh} or the baryon temperature $T_b$ is cooler than expected based on $\Lambda$CDM \citep{Bowman:2018yin,Barkana:2018lgd,Hill:2018lfx,Costa:2018aoy}. In the latter scenario, the EDGES measurement indicates that the baryon temperature $T_b$ at $z=20$ is  smaller than $7$K at 99\% C.L.

In Ref.~\citep{Hill:2018lfx}, the implications of EDGES were explored for an EDE model equivalent to the limit $n\to\infty$, including only the effect of EDE on the homogeneous evolution of densities and temperatures. Here we perform a similar analysis for $n=1$, $2$, and $3$, including perturbations in a ULA fluid.  

In the absence of any additional sources, the baryon gas temperature is driven by the balance between Compton heating and Hubble cooling
\be
    \frac{dT_b}{dz}
    = \frac{T_b(z) - T_{CMB}(z)}{(1+z) H(z) t_C(z)}
    + \frac{2T_b(z)}{(1+z)}\,.
    \label{eq:T_b(z)}
\ee
where $t_C(z)$ is the Compton-heating timescale. The key idea used in Ref.~\citep{Hill:2018lfx} is that, if the expansion rate before $z \sim 20$ is increased, the gas temperature decouples from the CMB temperature earlier, giving the gas more time to adiabatically cool. Within $\Lambda$CDM, baryons decouple around $z\sim 150$. To reach the 99\%~C.L. upper limit on the level of absorption measured by EDGES at $z\sim20$, the decoupling would need to happen around $z\sim 210$. The presence of a ULA that would dominate the expansion rate over a short period of time can potentially lead to a decoupling satisfying this condition. 

We show in Fig.~\ref{fig:Tb_ula} the baryon temperature at $z=20$ (close to the minimum of the absorption trough measured by EDGES \citep{Bowman:2018yin}) as a function of the ULA density today $\Omega_a^0$ and critical redshift $z_c$, for each value of $n$. To produce this figure, we fixed all $\Lambda$CDM parameters including the Hubble rate $H_0$ to values compatible with {\em Planck} 2015 data\footnote{In \texttt{CLASS}{}, these equations can be solved using either {\em Recfast} \citep{Seager:1999km,Seager:1999bc} or {\em HyRec} \citep{AliHaimoud:2010dx} and Eq.~(\ref{eq:T_b(z)}). Our choice of keeping $H_0$ fixed is motivated by the fact that adjusting $\theta_s$ requires strongly un-physical values of the Hubble rate (sometimes smaller than 0.01 km/s/Mpc) for which both {\em Recfast} and {\em HyRec} have difficulties to solve the cosmological recombination history. This also allows for a direct comparison with Ref.~\citep{Hill:2018lfx} where the same approach was used.}. Interestingly, we confirm that there exists a region of parameter space, centered around $z_c\sim100$ where the EDGES signal can be explained, in the $n=2$ and $n=3$ case. Our constraints on the ULA density from {\em Planck} data however strongly exclude all of these models, in agreement with Ref.~\citep{Hill:2018lfx}.

\section{Conclusions}
\label{sec:conclusions}

In this paper, we have studied the impact of ULAs on cosmological observations as they become dynamical at different times.  We have considered potentials of the form $V_n(\phi)\propto(1-\cos\phi)^n$, which show a wide variety of phenomenological consequences. At early times, each field is frozen in its potential due to Hubble friction such that their equations of state are dark-energy like, i.e. $w_a\simeq -1$. Once Hubble friction becomes weak enough, the field becomes dynamical and eventually starts to oscillate at the bottom of its potential. Once averaged over the oscillation period, the potential leads to an equation of state equal to $w_a\simeq (n-1)/(n+1)$. 

Such fields had been previously invoked in several contexts. First, ULAs with $n=1$ and becoming dynamical at early times ($z\gtrsim10^5$) are known to be a viable DM candidate. On the other hand, ULAs still frozen today are a viable dark energy candidate \citep{Hlozek:2014lca,Marsh:2015xka}. Second, a statistical ensemble of such fields may alleviate the coincidence problem today \citep{Griest:2002cu,Kamionkowski:2014zda,Emami:2016mrt,Karwal:2016vyq}. This general scenario may also provide a way to connect the physics of cosmic inflation to our current period of accelerated expansion \citep{Kamionkowski:2014zda}.  Third, the presence of an EDE can possibly reduce the Hubble constant tension \citep{Karwal:2016vyq} and explain the anomalously low baryon temperature inferred by the EDGES experiment \citep{Hill:2018lfx}. 

\begin{figure}[h!]
    \centering   \includegraphics[scale=0.35]{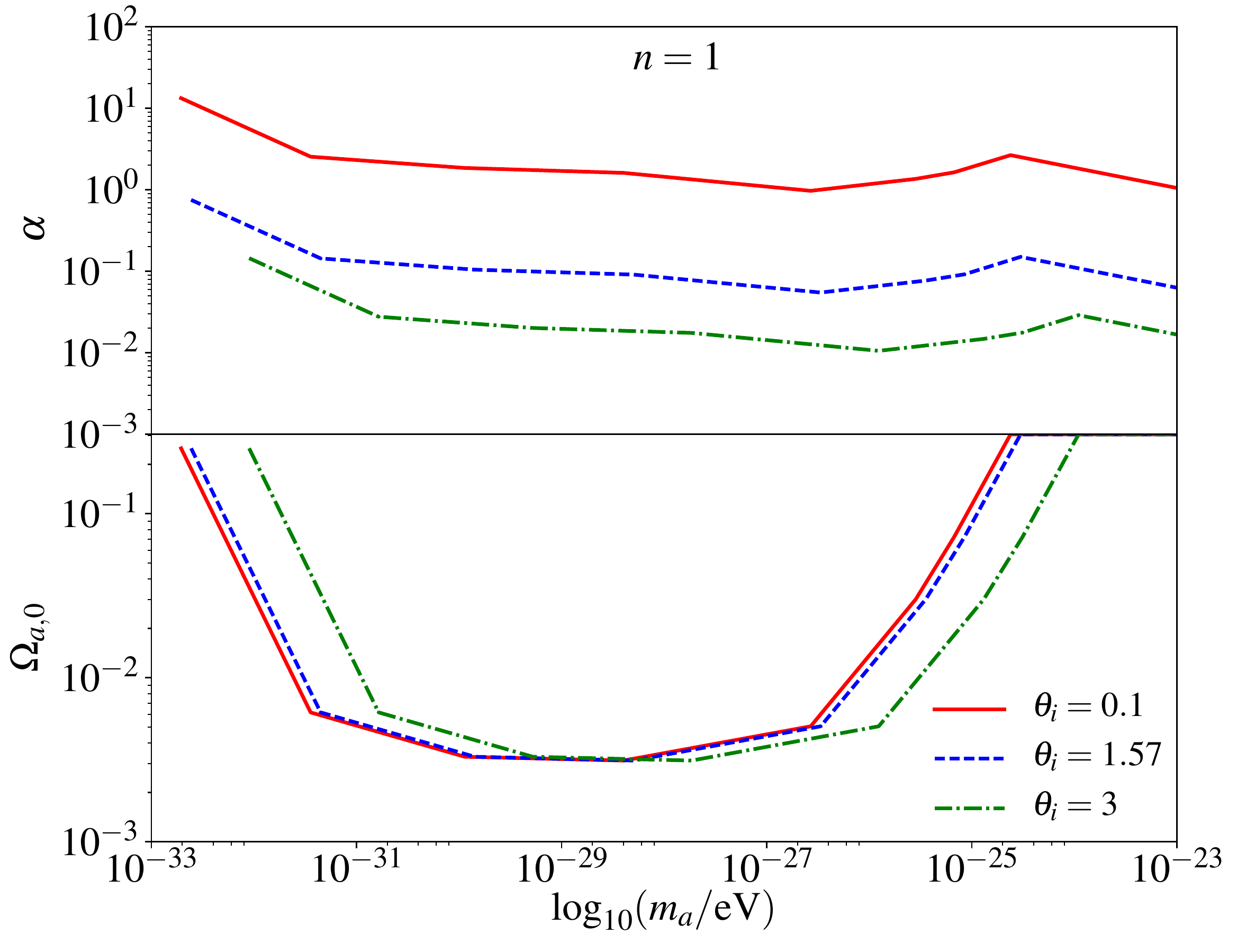}
    \includegraphics[scale=0.35]{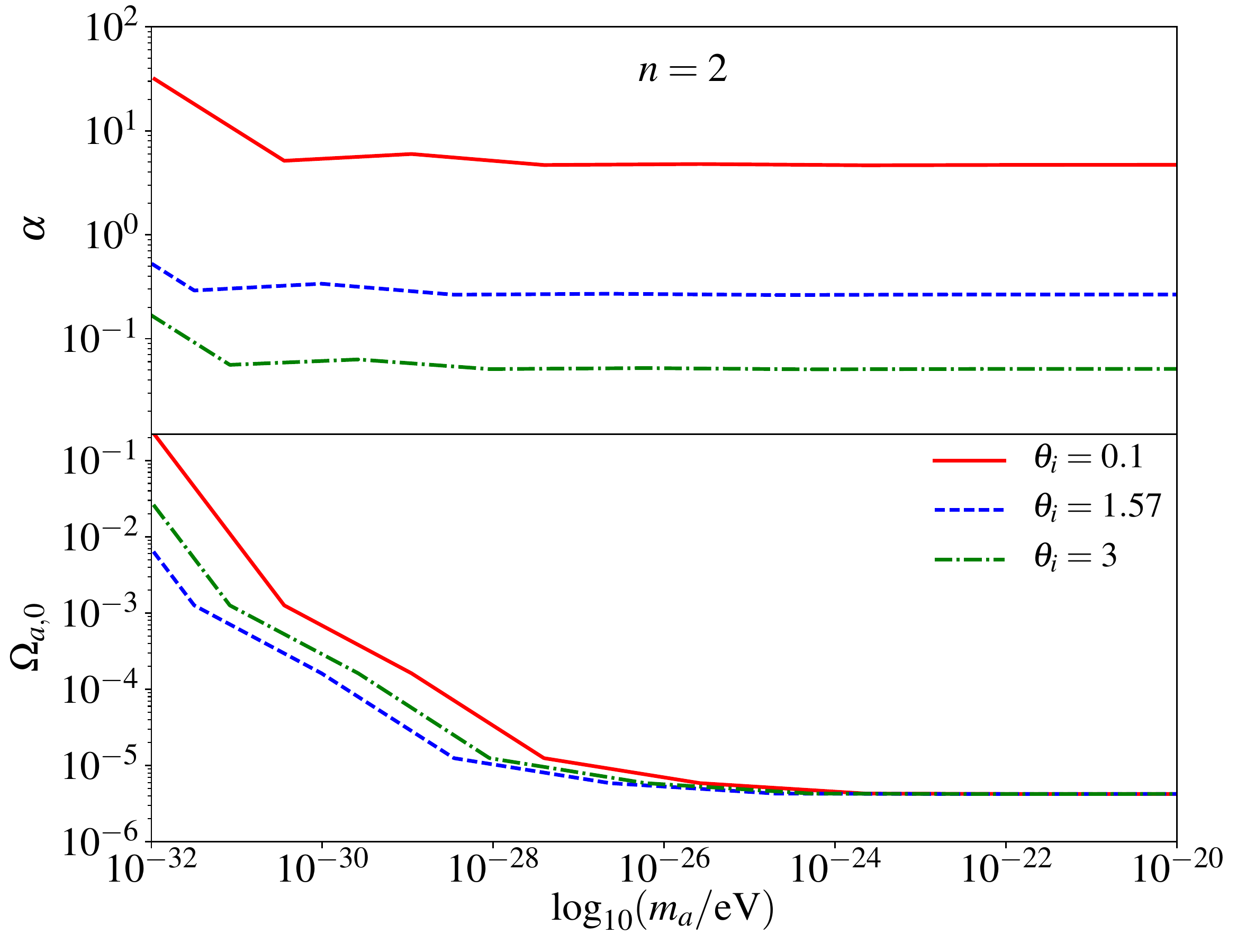}
    \includegraphics[scale=0.35]{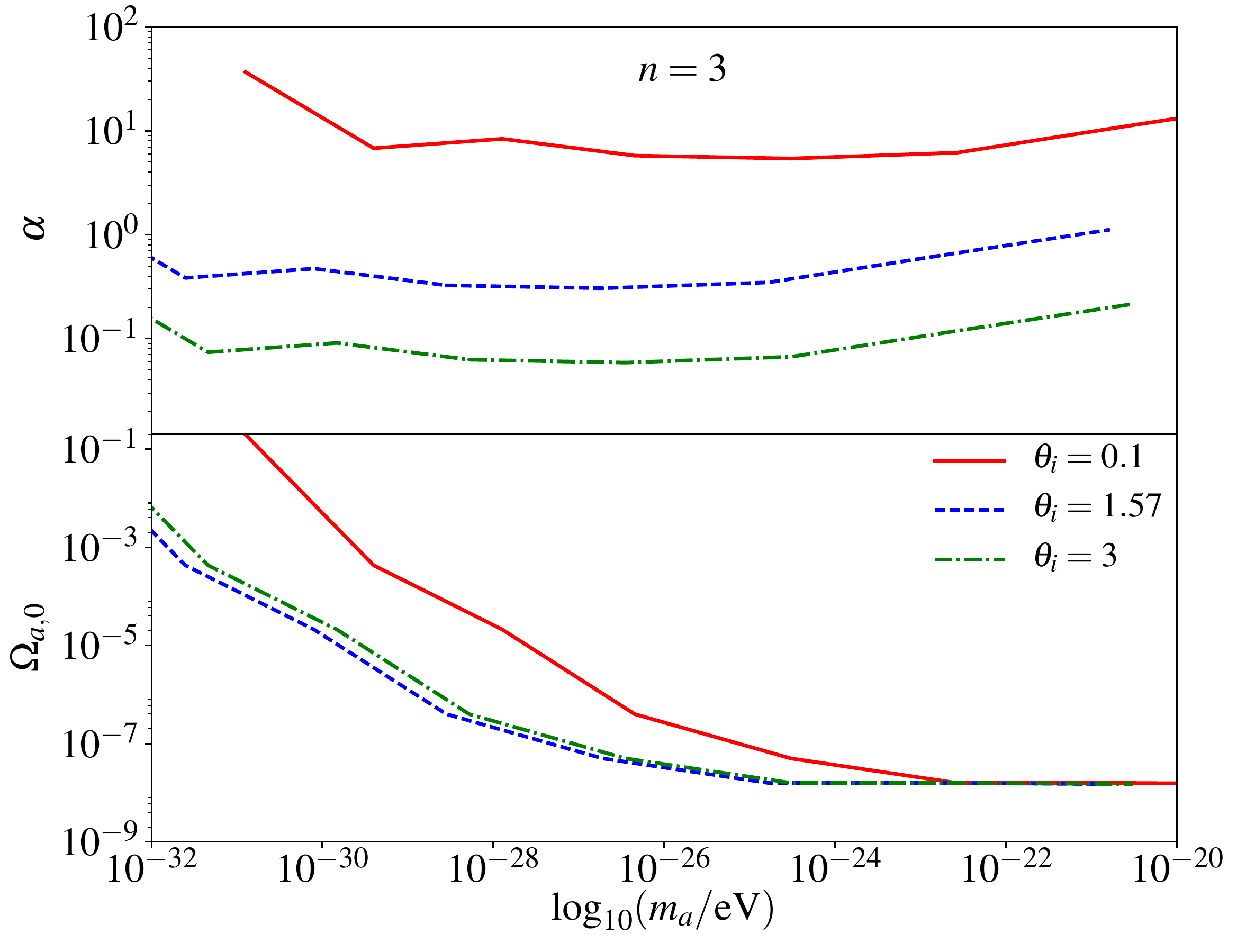}
    \caption{Constraints on the axion model parameters $(\mu, \alpha)$. We illustrate the impact of the initial field values by choosing $\theta_i = 0.1,\pi/2,3$.}
    \label{fig:constraints_axion_params}
\end{figure}

We have extended these previous studies in several significant ways. First, we have presented a parametrization of the ULA dynamics in terms of the redshift when the field becomes dynamical, $z_c$, and the fractional energy density in the axion field at $z_c$, $f_{z_c}$. Second of all, we have extended the effective fluid formalism for ULAs to anharmonic ULA potentials.  These perturbations can be approximately described by a time-averaged fluid component with a time and scale dependent effective sound speed \citep{Hu:2000ke,Hwang:2009js,Marsh:2010wq,Park:2012ru,Hlozek:2014lca,Marsh:2015xka,Noh:2017sdj} within the `generalized dark matter' parameterization \citep{Hu:1998kj}. Moreover, we derived a mapping between this parametrization and the ULA mass, decay constant and initial field value and attested of the accuracy of our fluid approximation by direct comparison with the exact KG solution in Appendix \ref{sec:app_full_vs_approx}. We have also shown that this WKB approximation is strictly only valid for potentials with $n\leq 3$, otherwise the period of oscillation is shorter than a Hubble time, violating the WKB assumptions.

Second, equipped with this fluid formalism we have compared the phenomenological consequences of axion-like potentials with $n=1$ which dilutes as cold dark matter (CDM), $n=2$ which dilutes as radiation, and $n=3$ which dilutes faster than radiation. We were thus able to explore any degeneracy the ULAs may have with known cosmological components, in particular CDM and neutrinos, and quantify the sensitivity of the data to a ULA component that decays even faster than radiation. We have constrained the abundance of ULAs as a function of $z_c$ using current cosmological data sets with a MCMC analysis, in order to fully explore degeneracies between the ULA parameters and the standard cosmological parameters. Remarkably, the details of the ULA effective sound speed could distinguish the effects of a ULA from other cosmological components, even if the ULA time-averaged equation of state is equal to zero (CDM-like) or 1/3 (radiation-like). Moreover, we have found that the CMB is sensitive to the field becoming dynamical as early as $z_c \sim 10^{10}$.

We illustrate in Fig.~\ref{fig:constraints_axion_params} how the constraints derived in this work in the $(z_c,\Omega_{a}^0)-$plane translate onto constraints on the axion parameters, i.e. ($\mu$,$\alpha$).
We use the relations introduced in Sec.~\ref{sec:translation} to map the ULA parameters to our fluid formalism. As an example, we choose three different initial field values, namely $\theta_i = 0.1,\pi/2,3$. The smallest value of $\theta_i$ allows direct comparison with results from Refs.~\citep{Hlozek:2017zzf,Hlozek:2014lca} (derived in the quadratic approximation), while the two others show how these constraints vary with the initial field value. We plan to study further implications of these constraints on the Axiverse in a forthcoming publication \citep{SmithXXX}.

Finally, we have studied the implications of our constraints for cosmological tensions. We have shown that fields with $n=2$ and $n=3$ can significantly ease the tension, as previously found for $n\to\infty$. However, our results put this scenario under strong pressure. On the one hand, the explanation of the EDGES signal is excluded by more than three orders of magnitude. On the other hand, we find that ULAs could at best ease the $H_0$ tension from $\sim3.4\sigma$ to $\sim 2\sigma$ given the level of our constraints. Contrary to expectation, the $n=2$ scenario is favored over $n=3$ even if the latter dilutes faster. This scenario also does slightly better than a relativistic species with arbitrary sound speed and viscosity, which can only relax the tension at the $2.4\sigma$ level \citep{Poulin:2018zxs}.

Our formalism represents a state-of-the-art treatment of the effect of ULAs on cosmological observables and can be used safely to analyse {\em Planck} data. In the future, CMB and LSS experiments with yet un-reached precision will be built. It is still to be established whether the fluid approximation will be accurate enough to describe the impact of ULAs without introducing strong bias in the reconstruction of cosmological parameters. However, this formalism is essential in order to perform extensive MCMC scan given the difficulty of solving the full KG equations. We plan to address the validity of our fluid formalism relative to solving the linear KG equations in light of the sensitivity of future experiments in forthcoming papers \citep{JonnyCAMB,AxiCLASS}.

\begin{acknowledgments}
This research project was conducted using computational resources
at the Maryland Advanced Research Computing
Center (MARCC). Part of this work has been done thanks to the facilities offered by the Universit\'{e} Savoie Mont-Blanc MUST computing center. V.~P.~and T~.K.~were supported at Johns Hopkins by NSF Grant No. 0244990, NASA NNX17AK38G, and the Simons Foundation.
T.~L.~S.~and D.~G.~acknowledge support in part by NASA ATP grant 17-ATP17-0162. T.~L.~S.~acknowledges support from the Provost's office at Swarthmore College. D.~G.~acknowledges support from the Provost's office at Haverford College. This work was supported in part by the National Science Foundation under Grant No. NSF PHY-1125915 at the Kavli Institute for Theoretical Physics (KITP) at UC Santa Barbara. We thank D.~J.~E.~Marsh for a thorough and insightful reading of the manuscript. We thank Charlotte Owen for her help in computing the exact ULA dynamics, as well as her very careful reading of the manuscript. D.~G.~thanks KITP for its hospitality during the completion of this work.  
\end{acknowledgments}
\begin{appendix}

\section{Derivation of the time-averaged effective sound speed for a generic oscillating potential} 
\label{sec:app_derivs}

Here we derive the effective sound speed, following the covariant perturbation theory notation used in Refs. \citep{Hu:1998kj,Hu:2004xd}. We can write the linearly perturbed Friedman-Robertson-Walker (FRW) metric as 
\begin{eqnarray}
g^{00} &=& -a^{-2}(1-2A),\\
g^{0i} &=& - a^{-2} B^i,\\
g^{ij} &=& a^{-2} (\gamma^{ij} - 2 H_L \gamma^{ij} - 2 H_T^{ij}),
\end{eqnarray}
where $\gamma_{ij} dx^i dx^j = d\chi^2 + \chi^2 d\Omega$ and $\chi$ is the comoving distance. Using conformal time, the equation of motion for the linear perturbation of the axion field is given by
\begin{eqnarray}
\ddot{\phi}_1 + 3 H \dot{\phi}_1 &+&\left(\frac{k^2}{a^2} + V''\right) \phi_1\label{eq:phi1EOM}\\ &=& (\dot A + 3 \dot H_L - k/a B) \dot{\phi}_0 - 2 A V',\nonumber
\end{eqnarray}
where $B$ is the longitudinal part of $B^i$.  In synchronous gauge we have $A=B=0$, $\eta \equiv -1/3 H_T - H_L$, $h\equiv 6 H_L$, where $\eta$ and $h$ are the metric variables used in Ref.~\citep{Ma:1995ey}.

We can write the density, pressure, and velocity perturbations in the scalar field stress energy as 
\begin{eqnarray}
\delta \rho_a &=& (\dot{\phi}_0 \dot{\phi}_1 - \dot{\phi}_0^2 A) + V' \phi_1, \\
\delta P_{a} &=& (\dot{\phi}_0 \dot{\phi}_1 - \dot{\phi}_0^2 A) - V' \phi_1,\\
T^0_i &=& \nabla_i Q_\phi =\nabla_i \dot{\phi}_0 \phi_1,
\end{eqnarray}
where $Q_\phi \equiv (\rho_{a}+p_{a}) (v_{a} - B)$.
We suppose that the field is oscillating about the minimum of its potential with a frequency $\varpi \gg H$ and we want to find the sound speed in the axion's average `rest frame'. In this rest frame when averaging over the fast oscillations we have
\begin{equation}
    \langle T_i^0 \rangle = 0 \rightarrow \langle \dot\phi_0 \phi_1 \rangle = 0,
\end{equation}
which fixes the gauge condition for the metric perturbation $B$.  We also require that in the axion rest-frame the time-averaged axion heat-flux is locally conserved:
\begin{equation}
\Bigg\langle \left[ \frac{\partial}{\partial \eta} + 4 \mathcal{H}\right] Q_a\Bigg\rangle = 0
\end{equation}
which, through the Euler equation for the axion stress energy, implies our second gauge condition \citep{Hu:2004xd,Hwang:2009js}
\begin{equation}
    \langle \rho_a + P_a \rangle A = - \langle \delta P_a \rangle. 
   \label{eq:gaugeA}
\end{equation}
We can write the linearly perturbed axion energy density as 
\begin{eqnarray}
    \delta \rho_a &=&  \dot \phi_0 \dot \phi_1 - (\rho_a + P_a) A + V' \phi_1,\\
    \delta P_a &=&  \dot \phi_0 \dot \phi_1 - (\rho_a + P_a) A - V' \phi_1,
\end{eqnarray}
which along with our gauge condition in Eq.~(\ref{eq:gaugeA}) gives 
\begin{eqnarray}
    \varpi^2 \langle  \phi_0  \phi_1 \rangle &=& \langle V' \phi_1 \rangle ,\\
    \langle \delta \rho_a\rangle  &=& \langle \delta P_a \rangle+2 \varpi^2 \langle \phi_0\phi_1\rangle  .
\end{eqnarray}

Keeping only the terms which vary on the (short) oscillation time-scale, the perturbed Klein-Gordon equation is: 
\begin{equation}
    \ddot{\phi}_1 + \left(\frac{k^2}{a^2}+ V''\right) \phi_1 \simeq -2 A V'.
\end{equation}
Multiplying this equation by $\phi_0$ and averaging over the short period  we have 
\begin{eqnarray}
  -\varpi^2 \langle \phi_1 \phi_0\rangle + \frac{k^2}{a^2}\langle \phi_1 \phi_0\rangle &+&(2n-1)  \langle V' \phi_1 \rangle\nonumber \\ &\simeq&- 4 A n\langle V \rangle
\end{eqnarray}
Finally, the virial theorem allows us to write 
\begin{eqnarray}
    \langle \rho_a \rangle &=& (n+1)  \langle V\rangle,\\
     \langle P_a \rangle &=&(n-1) \langle V\rangle,
\end{eqnarray}
so that 
\begin{equation}
    \langle \rho_a + P_a \rangle = 2n  \langle V\rangle,
\end{equation}
and the Klein-Gordon equation can be written 
\begin{eqnarray}
 \left( \frac{k^2}{a^2}+2 (n-1) \varpi ^2\right) \langle \phi_1 \phi_0\rangle  \simeq 2 \langle \delta P_a \rangle
\end{eqnarray}
This allows us to write 
\begin{eqnarray}
 c_s^2 \equiv \frac{\langle \delta P_a\rangle}{\langle \delta \rho_a\rangle} = \frac{2 a^2 (n-1) \varpi ^2+k^2}{2 a^2 (n+1) \varpi ^2+k^2}. \label{eq:APceff2}
\end{eqnarray}

The effective sound speed is computed in a gauge where 
\begin{eqnarray}
B &=& \langle v_a \rangle, \\
A &=& -\frac{\langle \delta P_a\rangle}{\langle \rho_a + P_{a}\rangle},
\end{eqnarray}
but we are doing our calculations in synchronous gauge where $A=B=0$. Next we will show that by transforming to synchronous gauge the effective sound speed enters into the fluid dynamics as dictated by the GDM equations of motion \citep{Hu:1998kj}.

A general gauge transformation takes the form 
\begin{eqnarray}
\eta &=& \tilde \eta + T,\\
x^i &=& \tilde x^i + L^i,
\end{eqnarray}
which leads to a transformation of the scalar metric potentials 
\begin{eqnarray}
A &=& \tilde A - \dot T - \mathcal{H} T,\\
B &=& \tilde B + \dot L + k T,\\
H_L &=& \tilde H_L - \frac{k}{3} L - \mathcal{H} T,\\
H_T &=& \tilde{H}_T + k L,
\end{eqnarray}
and transformation of the components of the stress-energy tensor
\begin{eqnarray}
\delta \rho_a &=& \delta \tilde \rho_{a} - \dot \rho_{a}T,\\
\delta P_{a} &=& \delta \tilde P_{a}- \dot  P_a T,\\
v_a &=& \tilde v_{a} + \dot L.
\end{eqnarray}
This tells us that to transform from our comoving gauge to synchronous gauge where $B=0$ we must have 
\begin{eqnarray}
\dot L + k T &=& - \langle v_a\rangle,
\end{eqnarray}
which in turn, using the transformation for the velocity, implies 
\begin{eqnarray}
T=-v_a/k,
\end{eqnarray}
where $v_\phi$ is the axion velocity perturbation in synchronous gauge. 

In order to determine how $c_s^2$ affects the evolution of the averaged field in synchronous gauge we now compute the synchronous gauge entropy perturbation, $P_a \Gamma_a \equiv \delta P_a - c_{\rm a}^2 \delta \rho_a$, where $c_{\rm a}^2 = \dot P_a/\dot \rho_a$, in terms of the averaged field variable in the comoving gauge. We start with an expression for the pressure perturbation in synchronous gauge:
\begin{eqnarray}
\delta P_a &=& \langle \delta P_a \rangle + \dot P_a v_a/k,\\
&=& c_s^2 \langle \delta \rho_a\rangle + \dot P_a v_a/k,
\end{eqnarray}
where in the second line we have used the effective sound speed. Next we write the comoving density perturbation in terms of the synchronous density perturbation and use the homogeneous continuity equation:
\begin{eqnarray}
\delta P_a &=& c_s^2 (\delta \rho_a - \dot\rho_a v_\phi/k)+\dot P_a v_a/k,\\
&=&c_s^2 \delta \rho_a +3 \mathcal{H}(1+w_a) \rho_a( c_s^2 - c_{\rm ad}^2)v_a/k.
\end{eqnarray}
This leads to 
\begin{equation}
P_a \Gamma_a = (c_s^2 - c_{\rm a}^2)\left[\delta \rho_a + 3 \mathcal{H}(1+w_a) \rho_a v_a/k\right].
\end{equation}
This implies that we can use the GDM equations of motion to approximate the evolution of the perturbations in the axion field with an effective sound-speed which transitions from $c_s^2 =1$ for $z> z_c$ to Eq.~(\ref{eq:APceff2}) for $z< z_c$.

\begin{figure}[h!]
    \includegraphics[scale=0.35]{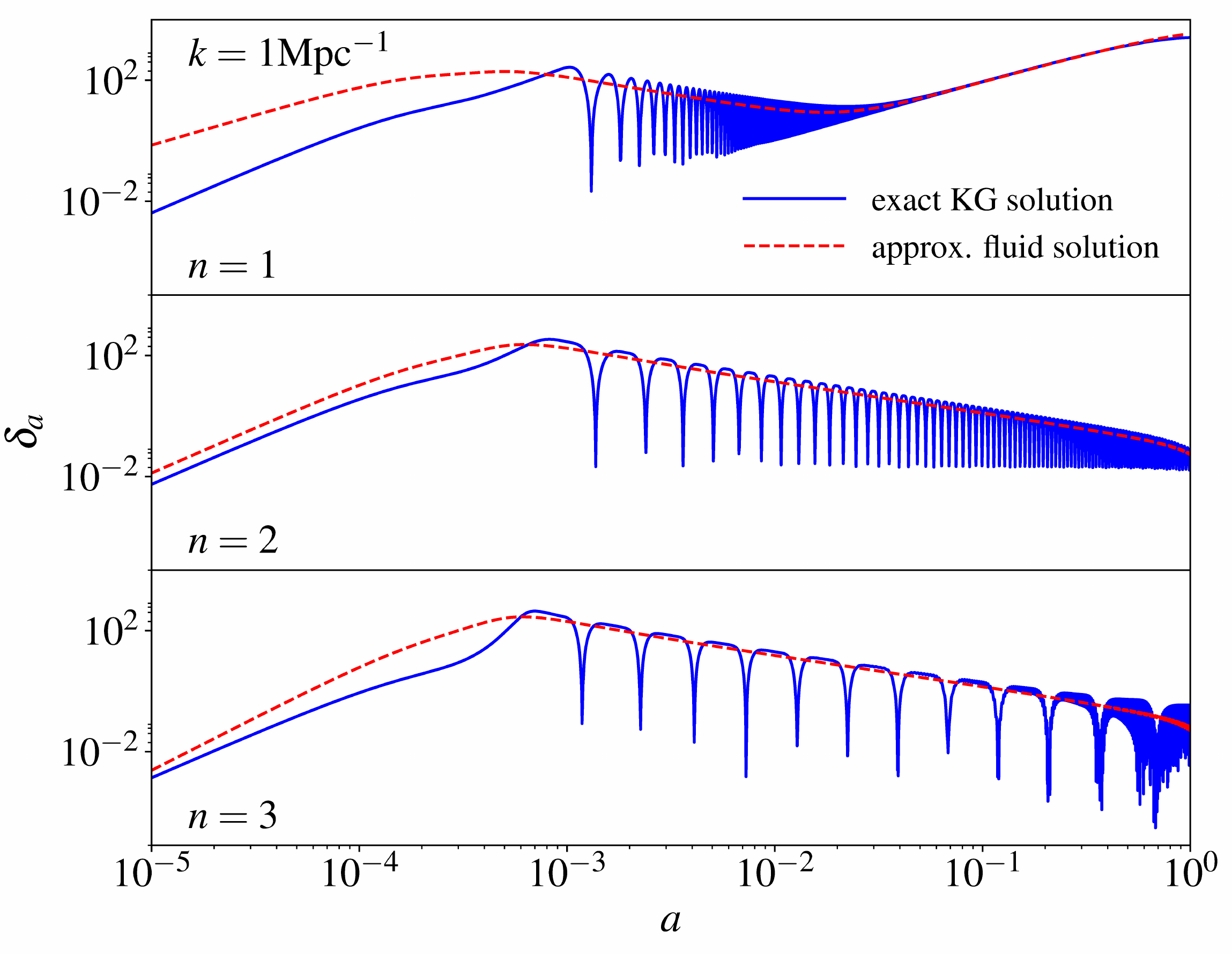}
    \includegraphics[scale=0.35]{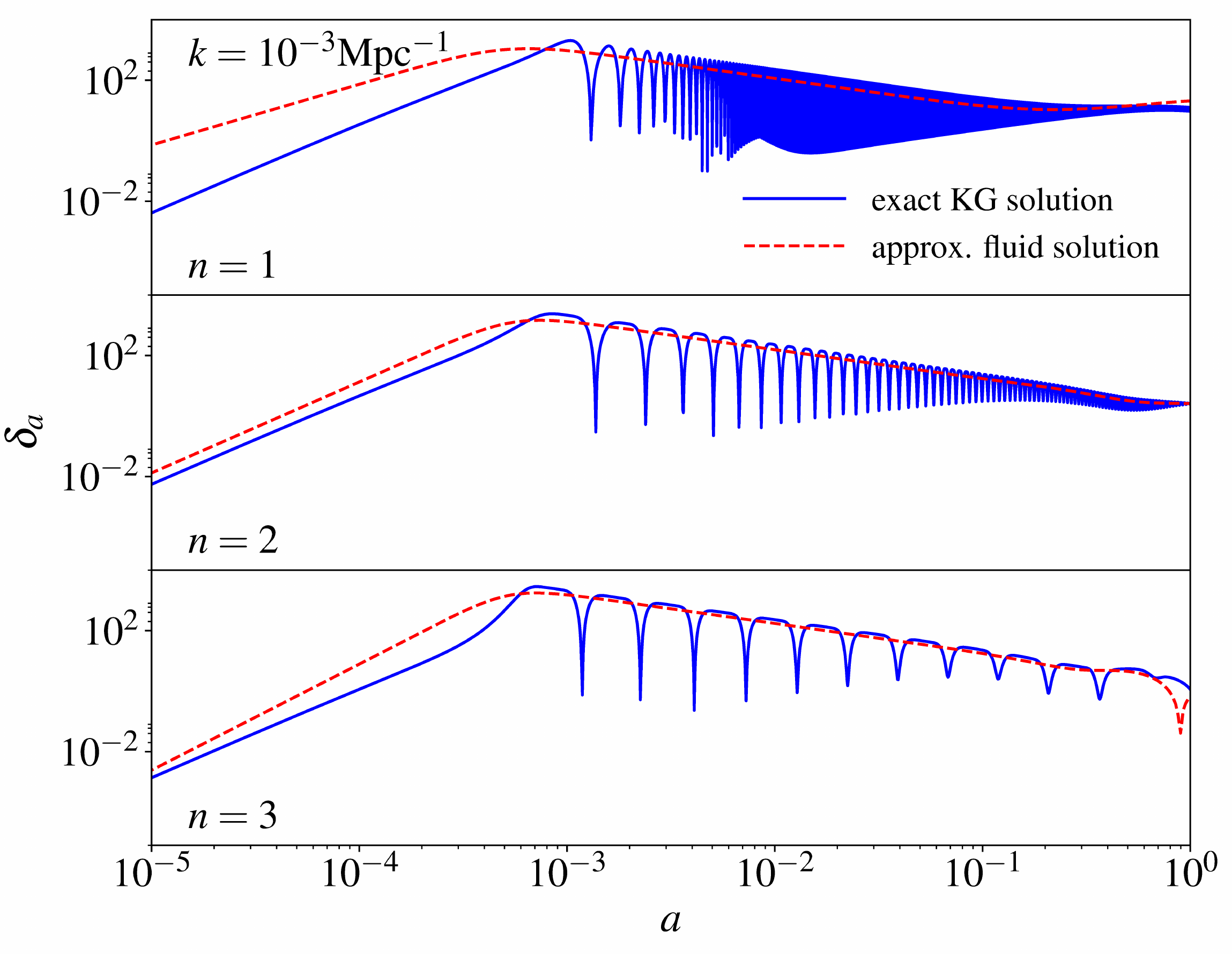}
    \caption{The exact and approximate evolution of density perturbations with wavenumber $k=1$ Mpc$^{-1}$ (top panel) and $k=10^{-3}$ Mpc$^{-1}$ (bottom panel) for  $n=1,2,3$ and $(\mu,\alpha)=(10^5,0.05)$.  The initial field values $\Theta_i$ were set to $1.5,2.5,3$, respectively.}
    \label{fig:perturb_full_vs_approx}
\end{figure}

\begin{figure}[t]
    \centering   \includegraphics[scale=0.35]{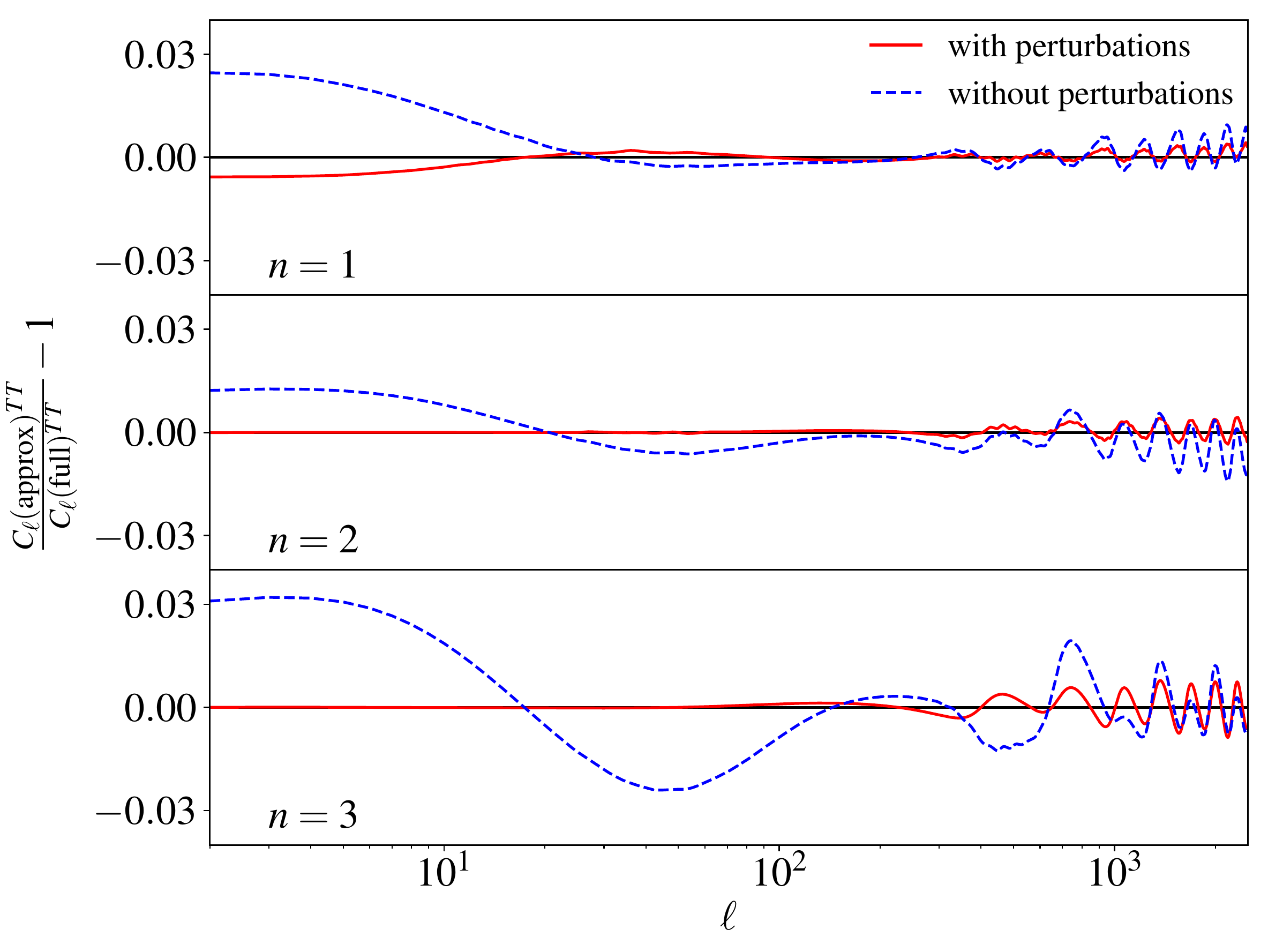}
    \includegraphics[scale=0.35]{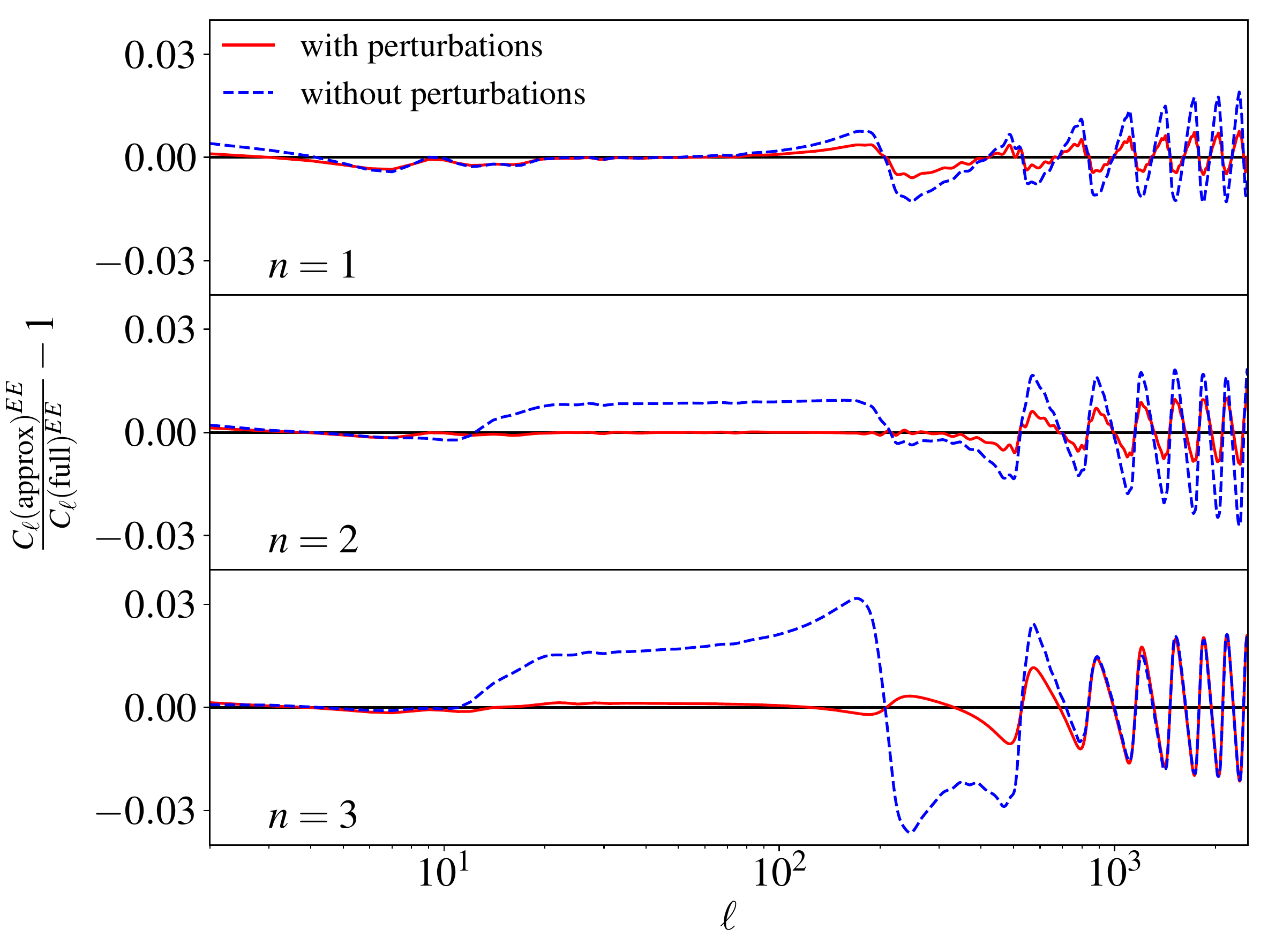}
    \caption{Residuals of the CMB TT (top panel) and EE (bottom panel) power spectra calculated in the fluid approximation with respect  to solving exactly the KG equations for $n=1,2,3$ and $(\mu,\alpha)=(10^5,0.05)$. The initial field values $\Theta_i$ were set to $1.5,2.5,3$. We show the case of neglecting perturbations of the ULA for comparison.   }
    \label{fig:powerspectra_full_vs_approx}
\end{figure}

\section{Approximate vs. exact dynamics}
\label{sec:app_full_vs_approx}

To check the validity of our fluid approach, we compare it to the solution of the full KG equations for specific (arbitrary) values $(\mu,\alpha)=(10^5,0.05)$. We choose an initial field value $\Theta_i= 1.5,2.5,3$ for $n = 1,2,3$ respectively. We use the relations introduced in Sec.~\ref{sec:translation} to map the ULA parameters to our fluid formalism. From a given $mu$, $\alpha$ and $\Theta_i$, we can easily calculate $\Omega_{a}(z_c)$ and $x_c$. We then make use of a shooting method in order to achieve eq.~\ref{eq:actoxc} (that cannot be solved analytically except if we assume that a single species dominate the universe energy content). We have checked that changing these parameters do not affect our conclusions. We plot the evolution of density perturbations with wavenumber $k=1,10^{-3}$ Mpc$^{-1}$ in Fig.~\ref{fig:perturb_full_vs_approx}. The impact of our approximation on the CMB and power spectra is shown in Fig.~\ref{fig:powerspectra_full_vs_approx}. We also show the case of neglecting perturbations of the ULA for comparison. 

By looking at Fig.~\ref{fig:perturb_full_vs_approx}, one can see that our parametrization captures well the overall behavior of the density perturbations. While it fails at following all of the oscillations, the envelope (i.e. the amplitude) of these is well reproduced. The agreement improves when the ULA starts oscillating, since our parametrization is designed for that regime. 

In Fig.~\ref{fig:powerspectra_full_vs_approx}, one can see that the CMB TT and EE power spectra are calculated at a few percent accuracy. The agreement is better for $n=1$ (it is always below a percent point) and degrades when going to higher power of $n$. This is expected as the WKB approximation, valid when the field oscillations are much more rapid than the Hubble time, breaks-down for $n \geq 3$. One can also gauge the impact of including perturbations: it is particularly important to avoid creating large deviations at multipoles $\ell \lesssim 100$. Remarkably, below multipoles of a few hundred the agreement is always well below a percent when including perturbations. Perturbations also have an impact at high multipoles, especially in the $n=1$ and 2 case, but does not improve the agreement very significantly in the $n=3$ case. 
From this quick comparison, we conclude that it is safe to use our parameterization given the precision of {\em Planck} data and the fact that we merely derive constraints on the ULAs abundances. However, we note that given the accuracy of next generation CMB experiments at high multipoles, searches for ULA in future cosmological data might require the evolution of the full KG equations (especially in the $n>1$ cases). We will investigate this possibility further in forthcoming publications \citep{JonnyCAMB,AxiCLASS}.

\bibliography{ULA.bib}

\begin{thebibliography}{82}
\expandafter\ifx\csname natexlab\endcsname\relax\def\natexlab#1{#1}\fi
\expandafter\ifx\csname bibnamefont\endcsname\relax
  \def\bibnamefont#1{#1}\fi
\expandafter\ifx\csname bibfnamefont\endcsname\relax
  \def\bibfnamefont#1{#1}\fi
\expandafter\ifx\csname citenamefont\endcsname\relax
  \def\citenamefont#1{#1}\fi
\expandafter\ifx\csname url\endcsname\relax
  \def\url#1{\texttt{#1}}\fi
\expandafter\ifx\csname urlprefix\endcsname\relax\def\urlprefix{URL }\fi
\providecommand{\bibinfo}[2]{#2}
\providecommand{\eprint}[2][]{\url{#2}}

\bibitem[{\citenamefont{Ade et~al.}(2016{\natexlab{a}})}]{Ade:2015xua}
\bibinfo{author}{\bibfnamefont{P.~A.~R.} \bibnamefont{Ade}}
  \bibnamefont{et~al.} (\bibinfo{collaboration}{Planck}),
  \bibinfo{journal}{Astron. Astrophys.} \textbf{\bibinfo{volume}{594}},
  \bibinfo{pages}{A13} (\bibinfo{year}{2016}{\natexlab{a}}),
  \eprint{1502.01589}.

\bibitem[{\citenamefont{Arvanitaki et~al.}(2010)\citenamefont{Arvanitaki,
  Dimopoulos, Dubovsky, Kaloper, and March-Russell}}]{Arvanitaki:2009fg}
\bibinfo{author}{\bibfnamefont{A.}~\bibnamefont{Arvanitaki}},
  \bibinfo{author}{\bibfnamefont{S.}~\bibnamefont{Dimopoulos}},
  \bibinfo{author}{\bibfnamefont{S.}~\bibnamefont{Dubovsky}},
  \bibinfo{author}{\bibfnamefont{N.}~\bibnamefont{Kaloper}}, \bibnamefont{and}
  \bibinfo{author}{\bibfnamefont{J.}~\bibnamefont{March-Russell}},
  \bibinfo{journal}{Phys. Rev.} \textbf{\bibinfo{volume}{D81}},
  \bibinfo{pages}{123530} (\bibinfo{year}{2010}), \eprint{0905.4720}.

\bibitem[{\citenamefont{Marsh}(2016)}]{Marsh:2015xka}
\bibinfo{author}{\bibfnamefont{D.~J.~E.} \bibnamefont{Marsh}},
  \bibinfo{journal}{Phys. Rept.} \textbf{\bibinfo{volume}{643}},
  \bibinfo{pages}{1} (\bibinfo{year}{2016}), \eprint{1510.07633}.

\bibitem[{\citenamefont{Karwal and Kamionkowski}(2016)}]{Karwal:2016vyq}
\bibinfo{author}{\bibfnamefont{T.}~\bibnamefont{Karwal}} \bibnamefont{and}
  \bibinfo{author}{\bibfnamefont{M.}~\bibnamefont{Kamionkowski}},
  \bibinfo{journal}{Phys. Rev.} \textbf{\bibinfo{volume}{D94}},
  \bibinfo{pages}{103523} (\bibinfo{year}{2016}), \eprint{1608.01309}.

\bibitem[{\citenamefont{Hill and Baxter}(2018)}]{Hill:2018lfx}
\bibinfo{author}{\bibfnamefont{J.~C.} \bibnamefont{Hill}} \bibnamefont{and}
  \bibinfo{author}{\bibfnamefont{E.~J.} \bibnamefont{Baxter}}
  (\bibinfo{year}{2018}), \eprint{1803.07555}.

\bibitem[{\citenamefont{Feeney et~al.}(2017)\citenamefont{Feeney, Mortlock, and
  Dalmasso}}]{Feeney:2017sgx}
\bibinfo{author}{\bibfnamefont{S.~M.} \bibnamefont{Feeney}},
  \bibinfo{author}{\bibfnamefont{D.~J.} \bibnamefont{Mortlock}},
  \bibnamefont{and} \bibinfo{author}{\bibfnamefont{N.}~\bibnamefont{Dalmasso}}
  (\bibinfo{year}{2017}), \eprint{1707.00007}.

\bibitem[{\citenamefont{Riess et~al.}(2016)}]{Riess:2016jrr}
\bibinfo{author}{\bibfnamefont{A.~G.} \bibnamefont{Riess}}
  \bibnamefont{et~al.}, \bibinfo{journal}{Astrophys. J.}
  \textbf{\bibinfo{volume}{826}}, \bibinfo{pages}{56} (\bibinfo{year}{2016}),
  \eprint{1604.01424}.

\bibitem[{\citenamefont{Bowman et~al.}(2018)\citenamefont{Bowman, Rogers,
  Monsalve, Mozdzen, and Mahesh}}]{Bowman:2018yin}
\bibinfo{author}{\bibfnamefont{J.~D.} \bibnamefont{Bowman}},
  \bibinfo{author}{\bibfnamefont{A.~E.~E.} \bibnamefont{Rogers}},
  \bibinfo{author}{\bibfnamefont{R.~A.} \bibnamefont{Monsalve}},
  \bibinfo{author}{\bibfnamefont{T.~J.} \bibnamefont{Mozdzen}},
  \bibnamefont{and} \bibinfo{author}{\bibfnamefont{N.}~\bibnamefont{Mahesh}},
  \bibinfo{journal}{Nature} \textbf{\bibinfo{volume}{555}}, \bibinfo{pages}{67}
  (\bibinfo{year}{2018}).

\bibitem[{\citenamefont{Hills et~al.}(2018)\citenamefont{Hills, Kulkarni,
  Meerburg, and Puchwein}}]{Hills:2018vyr}
\bibinfo{author}{\bibfnamefont{R.}~\bibnamefont{Hills}},
  \bibinfo{author}{\bibfnamefont{G.}~\bibnamefont{Kulkarni}},
  \bibinfo{author}{\bibfnamefont{P.~D.} \bibnamefont{Meerburg}},
  \bibnamefont{and} \bibinfo{author}{\bibfnamefont{E.}~\bibnamefont{Puchwein}}
  (\bibinfo{year}{2018}), \eprint{1805.01421}.

\bibitem[{\citenamefont{Bozek et~al.}(2015)\citenamefont{Bozek, Marsh, Silk,
  and Wyse}}]{Bozek:2014uqa}
\bibinfo{author}{\bibfnamefont{B.}~\bibnamefont{Bozek}},
  \bibinfo{author}{\bibfnamefont{D.~J.~E.} \bibnamefont{Marsh}},
  \bibinfo{author}{\bibfnamefont{J.}~\bibnamefont{Silk}}, \bibnamefont{and}
  \bibinfo{author}{\bibfnamefont{R.~F.~G.} \bibnamefont{Wyse}},
  \bibinfo{journal}{Mon. Not. Roy. Astron. Soc.}
  \textbf{\bibinfo{volume}{450}}, \bibinfo{pages}{209} (\bibinfo{year}{2015}),
  \eprint{1409.3544}.

\bibitem[{\citenamefont{Lidz and Hui}(2018)}]{Lidz:2018fqo}
\bibinfo{author}{\bibfnamefont{A.}~\bibnamefont{Lidz}} \bibnamefont{and}
  \bibinfo{author}{\bibfnamefont{L.}~\bibnamefont{Hui}} (\bibinfo{year}{2018}),
  \eprint{1805.01253}.

\bibitem[{\citenamefont{Svrcek and Witten}(2006)}]{Svrcek:2006yi}
\bibinfo{author}{\bibfnamefont{P.}~\bibnamefont{Svrcek}} \bibnamefont{and}
  \bibinfo{author}{\bibfnamefont{E.}~\bibnamefont{Witten}},
  \bibinfo{journal}{JHEP} \textbf{\bibinfo{volume}{06}}, \bibinfo{pages}{051}
  (\bibinfo{year}{2006}), \eprint{hep-th/0605206}.

\bibitem[{\citenamefont{Stott et~al.}(2017)\citenamefont{Stott, Marsh,
  Pongkitivanichkul, Price, and Acharya}}]{Stott:2017hvl}
\bibinfo{author}{\bibfnamefont{M.~J.} \bibnamefont{Stott}},
  \bibinfo{author}{\bibfnamefont{D.~J.~E.} \bibnamefont{Marsh}},
  \bibinfo{author}{\bibfnamefont{C.}~\bibnamefont{Pongkitivanichkul}},
  \bibinfo{author}{\bibfnamefont{L.~C.} \bibnamefont{Price}}, \bibnamefont{and}
  \bibinfo{author}{\bibfnamefont{B.~S.} \bibnamefont{Acharya}},
  \bibinfo{journal}{Phys. Rev.} \textbf{\bibinfo{volume}{D96}},
  \bibinfo{pages}{083510} (\bibinfo{year}{2017}), \eprint{1706.03236}.

\bibitem[{\citenamefont{Hlozek et~al.}(2015)\citenamefont{Hlozek, Grin, Marsh,
  and Ferreira}}]{Hlozek:2014lca}
\bibinfo{author}{\bibfnamefont{R.}~\bibnamefont{Hlozek}},
  \bibinfo{author}{\bibfnamefont{D.}~\bibnamefont{Grin}},
  \bibinfo{author}{\bibfnamefont{D.~J.~E.} \bibnamefont{Marsh}},
  \bibnamefont{and} \bibinfo{author}{\bibfnamefont{P.~G.}
  \bibnamefont{Ferreira}}, \bibinfo{journal}{Phys. Rev.}
  \textbf{\bibinfo{volume}{D91}}, \bibinfo{pages}{103512}
  (\bibinfo{year}{2015}), \eprint{1410.2896}.

\bibitem[{\citenamefont{Mukherjee et~al.}(2018)\citenamefont{Mukherjee, Khatri,
  and Wandelt}}]{Mukherjee:2018oeb}
\bibinfo{author}{\bibfnamefont{S.}~\bibnamefont{Mukherjee}},
  \bibinfo{author}{\bibfnamefont{R.}~\bibnamefont{Khatri}}, \bibnamefont{and}
  \bibinfo{author}{\bibfnamefont{B.~D.} \bibnamefont{Wandelt}},
  \bibinfo{journal}{JCAP} \textbf{\bibinfo{volume}{1804}}, \bibinfo{pages}{045}
  (\bibinfo{year}{2018}), \eprint{1801.09701}.

\bibitem[{\citenamefont{Kappl et~al.}(2016)\citenamefont{Kappl, Nilles, and
  Winkler}}]{Kappl:2015esy}
\bibinfo{author}{\bibfnamefont{R.}~\bibnamefont{Kappl}},
  \bibinfo{author}{\bibfnamefont{H.~P.} \bibnamefont{Nilles}},
  \bibnamefont{and} \bibinfo{author}{\bibfnamefont{M.~W.}
  \bibnamefont{Winkler}}, \bibinfo{journal}{Phys. Lett.}
  \textbf{\bibinfo{volume}{B753}}, \bibinfo{pages}{653} (\bibinfo{year}{2016}),
  \eprint{1511.05560}.

\bibitem[{\citenamefont{Turner}(1983)}]{PhysRevD.28.1243}
\bibinfo{author}{\bibfnamefont{M.~S.} \bibnamefont{Turner}},
  \bibinfo{journal}{Phys. Rev. D} \textbf{\bibinfo{volume}{28}},
  \bibinfo{pages}{1243} (\bibinfo{year}{1983}),
  \urlprefix\url{https://link.aps.org/doi/10.1103/PhysRevD.28.1243}.

\bibitem[{\citenamefont{Griest}(2002)}]{Griest:2002cu}
\bibinfo{author}{\bibfnamefont{K.}~\bibnamefont{Griest}},
  \bibinfo{journal}{Phys. Rev.} \textbf{\bibinfo{volume}{D66}},
  \bibinfo{pages}{123501} (\bibinfo{year}{2002}), \eprint{astro-ph/0202052}.

\bibitem[{\citenamefont{Linder and Smith}(2011)}]{Linder:2010wp}
\bibinfo{author}{\bibfnamefont{E.~V.} \bibnamefont{Linder}} \bibnamefont{and}
  \bibinfo{author}{\bibfnamefont{T.~L.} \bibnamefont{Smith}},
  \bibinfo{journal}{JCAP} \textbf{\bibinfo{volume}{1104}}, \bibinfo{pages}{001}
  (\bibinfo{year}{2011}), \eprint{1009.3500}.

\bibitem[{\citenamefont{Kamionkowski et~al.}(2014)\citenamefont{Kamionkowski,
  Pradler, and Walker}}]{Kamionkowski:2014zda}
\bibinfo{author}{\bibfnamefont{M.}~\bibnamefont{Kamionkowski}},
  \bibinfo{author}{\bibfnamefont{J.}~\bibnamefont{Pradler}}, \bibnamefont{and}
  \bibinfo{author}{\bibfnamefont{D.~G.~E.} \bibnamefont{Walker}},
  \bibinfo{journal}{Phys. Rev. Lett.} \textbf{\bibinfo{volume}{113}},
  \bibinfo{pages}{251302} (\bibinfo{year}{2014}), \eprint{1409.0549}.

\bibitem[{\citenamefont{Emami et~al.}(2016)\citenamefont{Emami, Grin, Pradler,
  Raccanelli, and Kamionkowski}}]{Emami:2016mrt}
\bibinfo{author}{\bibfnamefont{R.}~\bibnamefont{Emami}},
  \bibinfo{author}{\bibfnamefont{D.}~\bibnamefont{Grin}},
  \bibinfo{author}{\bibfnamefont{J.}~\bibnamefont{Pradler}},
  \bibinfo{author}{\bibfnamefont{A.}~\bibnamefont{Raccanelli}},
  \bibnamefont{and}
  \bibinfo{author}{\bibfnamefont{M.}~\bibnamefont{Kamionkowski}},
  \bibinfo{journal}{Phys. Rev.} \textbf{\bibinfo{volume}{D93}},
  \bibinfo{pages}{123005} (\bibinfo{year}{2016}), \eprint{1603.04851}.

\bibitem[{\citenamefont{Hu et~al.}(2000)\citenamefont{Hu, Barkana, and
  Gruzinov}}]{Hu:2000ke}
\bibinfo{author}{\bibfnamefont{W.}~\bibnamefont{Hu}},
  \bibinfo{author}{\bibfnamefont{R.}~\bibnamefont{Barkana}}, \bibnamefont{and}
  \bibinfo{author}{\bibfnamefont{A.}~\bibnamefont{Gruzinov}},
  \bibinfo{journal}{Phys. Rev. Lett.} \textbf{\bibinfo{volume}{85}},
  \bibinfo{pages}{1158} (\bibinfo{year}{2000}), \eprint{astro-ph/0003365}.

\bibitem[{\citenamefont{Hwang and Noh}(2009)}]{Hwang:2009js}
\bibinfo{author}{\bibfnamefont{J.-c.} \bibnamefont{Hwang}} \bibnamefont{and}
  \bibinfo{author}{\bibfnamefont{H.}~\bibnamefont{Noh}},
  \bibinfo{journal}{Phys. Lett.} \textbf{\bibinfo{volume}{B680}},
  \bibinfo{pages}{1} (\bibinfo{year}{2009}), \eprint{0902.4738}.

\bibitem[{\citenamefont{Marsh and Ferreira}(2010)}]{Marsh:2010wq}
\bibinfo{author}{\bibfnamefont{D.~J.~E.} \bibnamefont{Marsh}} \bibnamefont{and}
  \bibinfo{author}{\bibfnamefont{P.~G.} \bibnamefont{Ferreira}},
  \bibinfo{journal}{Phys. Rev.} \textbf{\bibinfo{volume}{D82}},
  \bibinfo{pages}{103528} (\bibinfo{year}{2010}), \eprint{1009.3501}.

\bibitem[{\citenamefont{Park et~al.}(2012)\citenamefont{Park, Hwang, and
  Noh}}]{Park:2012ru}
\bibinfo{author}{\bibfnamefont{C.-G.} \bibnamefont{Park}},
  \bibinfo{author}{\bibfnamefont{J.-c.} \bibnamefont{Hwang}}, \bibnamefont{and}
  \bibinfo{author}{\bibfnamefont{H.}~\bibnamefont{Noh}},
  \bibinfo{journal}{Phys. Rev.} \textbf{\bibinfo{volume}{D86}},
  \bibinfo{pages}{083535} (\bibinfo{year}{2012}), \eprint{1207.3124}.

\bibitem[{\citenamefont{Noh et~al.}(2017)\citenamefont{Noh, Hwang, and
  Park}}]{Noh:2017sdj}
\bibinfo{author}{\bibfnamefont{H.}~\bibnamefont{Noh}},
  \bibinfo{author}{\bibfnamefont{J.-c.} \bibnamefont{Hwang}}, \bibnamefont{and}
  \bibinfo{author}{\bibfnamefont{C.-G.} \bibnamefont{Park}},
  \bibinfo{journal}{Astrophys. J.} \textbf{\bibinfo{volume}{846}},
  \bibinfo{pages}{1} (\bibinfo{year}{2017}), \eprint{1707.08568}.

\bibitem[{\citenamefont{Hu}(1998)}]{Hu:1998kj}
\bibinfo{author}{\bibfnamefont{W.}~\bibnamefont{Hu}},
  \bibinfo{journal}{Astrophys. J.} \textbf{\bibinfo{volume}{506}},
  \bibinfo{pages}{485} (\bibinfo{year}{1998}), \eprint{astro-ph/9801234}.

\bibitem[{\citenamefont{Lyth}(1992)}]{Lyth:1991ub}
\bibinfo{author}{\bibfnamefont{D.~H.} \bibnamefont{Lyth}},
  \bibinfo{journal}{Phys. Rev.} \textbf{\bibinfo{volume}{D45}},
  \bibinfo{pages}{3394} (\bibinfo{year}{1992}).

\bibitem[{\citenamefont{Cembranos et~al.}(2016)\citenamefont{Cembranos, Maroto,
  and Núñez~Jareño}}]{Cembranos:2015oya}
\bibinfo{author}{\bibfnamefont{J.~A.~R.} \bibnamefont{Cembranos}},
  \bibinfo{author}{\bibfnamefont{A.~L.} \bibnamefont{Maroto}},
  \bibnamefont{and} \bibinfo{author}{\bibfnamefont{S.~J.}
  \bibnamefont{Núñez~Jareño}}, \bibinfo{journal}{JHEP}
  \textbf{\bibinfo{volume}{03}}, \bibinfo{pages}{013} (\bibinfo{year}{2016}),
  \eprint{1509.08819}.

\bibitem[{\citenamefont{Fan}(2016)}]{Fan:2016rda}
\bibinfo{author}{\bibfnamefont{J.}~\bibnamefont{Fan}}, \bibinfo{journal}{Phys.
  Dark Univ.} \textbf{\bibinfo{volume}{14}}, \bibinfo{pages}{84}
  (\bibinfo{year}{2016}), \eprint{1603.06580}.

\bibitem[{\citenamefont{Betoule et~al.}(2014)}]{Betoule:2014frx}
\bibinfo{author}{\bibfnamefont{M.}~\bibnamefont{Betoule}} \bibnamefont{et~al.}
  (\bibinfo{collaboration}{SDSS}), \bibinfo{journal}{Astron. Astrophys.}
  \textbf{\bibinfo{volume}{568}}, \bibinfo{pages}{A22} (\bibinfo{year}{2014}),
  \eprint{1401.4064}.

\bibitem[{\citenamefont{Lesgourgues}(2011{\natexlab{a}})}]{Lesgourgues:2011re}
\bibinfo{author}{\bibfnamefont{J.}~\bibnamefont{Lesgourgues}}
  (\bibinfo{year}{2011}{\natexlab{a}}), \eprint{1104.2932}.

\bibitem[{\citenamefont{Blas et~al.}(2011)\citenamefont{Blas, Lesgourgues, and
  Tram}}]{Blas:2011rf}
\bibinfo{author}{\bibfnamefont{D.}~\bibnamefont{Blas}},
  \bibinfo{author}{\bibfnamefont{J.}~\bibnamefont{Lesgourgues}},
  \bibnamefont{and} \bibinfo{author}{\bibfnamefont{T.}~\bibnamefont{Tram}},
  \bibinfo{journal}{JCAP} \textbf{\bibinfo{volume}{1107}}, \bibinfo{pages}{034}
  (\bibinfo{year}{2011}), \eprint{1104.2933}.

\bibitem[{\citenamefont{Lesgourgues}(2011{\natexlab{b}})}]{Lesgourgues:2011rg}
\bibinfo{author}{\bibfnamefont{J.}~\bibnamefont{Lesgourgues}}
  (\bibinfo{year}{2011}{\natexlab{b}}), \eprint{1104.2934}.

\bibitem[{\citenamefont{Lesgourgues and Tram}(2011)}]{Lesgourgues:2011rh}
\bibinfo{author}{\bibfnamefont{J.}~\bibnamefont{Lesgourgues}} \bibnamefont{and}
  \bibinfo{author}{\bibfnamefont{T.}~\bibnamefont{Tram}},
  \bibinfo{journal}{JCAP} \textbf{\bibinfo{volume}{1109}}, \bibinfo{pages}{032}
  (\bibinfo{year}{2011}), \eprint{1104.2935}.

\bibitem[{\citenamefont{Audren et~al.}(2013)\citenamefont{Audren, Lesgourgues,
  Benabed, and Prunet}}]{Audren:2012wb}
\bibinfo{author}{\bibfnamefont{B.}~\bibnamefont{Audren}},
  \bibinfo{author}{\bibfnamefont{J.}~\bibnamefont{Lesgourgues}},
  \bibinfo{author}{\bibfnamefont{K.}~\bibnamefont{Benabed}}, \bibnamefont{and}
  \bibinfo{author}{\bibfnamefont{S.}~\bibnamefont{Prunet}},
  \bibinfo{journal}{JCAP} \textbf{\bibinfo{volume}{1302}}, \bibinfo{pages}{001}
  (\bibinfo{year}{2013}), \eprint{1210.7183}.

\bibitem[{\citenamefont{Li et~al.}(2014)\citenamefont{Li, Rindler-Daller, and
  Shapiro}}]{Li:2013nal}
\bibinfo{author}{\bibfnamefont{B.}~\bibnamefont{Li}},
  \bibinfo{author}{\bibfnamefont{T.}~\bibnamefont{Rindler-Daller}},
  \bibnamefont{and} \bibinfo{author}{\bibfnamefont{P.~R.}
  \bibnamefont{Shapiro}}, \bibinfo{journal}{Phys. Rev.}
  \textbf{\bibinfo{volume}{D89}}, \bibinfo{pages}{083536}
  (\bibinfo{year}{2014}), \eprint{1310.6061}.

\bibitem[{\citenamefont{Li et~al.}(2017)\citenamefont{Li, Shapiro, and
  Rindler-Daller}}]{Li:2016mmc}
\bibinfo{author}{\bibfnamefont{B.}~\bibnamefont{Li}},
  \bibinfo{author}{\bibfnamefont{P.~R.} \bibnamefont{Shapiro}},
  \bibnamefont{and}
  \bibinfo{author}{\bibfnamefont{T.}~\bibnamefont{Rindler-Daller}},
  \bibinfo{journal}{Phys. Rev.} \textbf{\bibinfo{volume}{D96}},
  \bibinfo{pages}{063505} (\bibinfo{year}{2017}), \eprint{1611.07961}.

\bibitem[{\citenamefont{Johnson and Kamionkowski}(2008)}]{Johnson:2008se}
\bibinfo{author}{\bibfnamefont{M.~C.} \bibnamefont{Johnson}} \bibnamefont{and}
  \bibinfo{author}{\bibfnamefont{M.}~\bibnamefont{Kamionkowski}},
  \bibinfo{journal}{Phys. Rev.} \textbf{\bibinfo{volume}{D78}},
  \bibinfo{pages}{063010} (\bibinfo{year}{2008}), \eprint{0805.1748}.

\bibitem[{\citenamefont{Ma and Bertschinger}(1995)}]{Ma:1995ey}
\bibinfo{author}{\bibfnamefont{C.-P.} \bibnamefont{Ma}} \bibnamefont{and}
  \bibinfo{author}{\bibfnamefont{E.}~\bibnamefont{Bertschinger}},
  \bibinfo{journal}{Astrophys. J.} \textbf{\bibinfo{volume}{455}},
  \bibinfo{pages}{7} (\bibinfo{year}{1995}), \eprint{astro-ph/9506072}.

\bibitem[{\citenamefont{Ballesteros and
  Lesgourgues}(2010)}]{Ballesteros:2010ks}
\bibinfo{author}{\bibfnamefont{G.}~\bibnamefont{Ballesteros}} \bibnamefont{and}
  \bibinfo{author}{\bibfnamefont{J.}~\bibnamefont{Lesgourgues}},
  \bibinfo{journal}{JCAP} \textbf{\bibinfo{volume}{1010}}, \bibinfo{pages}{014}
  (\bibinfo{year}{2010}), \eprint{1004.5509}.

\bibitem[{\citenamefont{Hlozek et~al.}(2018)\citenamefont{Hlozek, Marsh, and
  Grin}}]{Hlozek:2017zzf}
\bibinfo{author}{\bibfnamefont{R.}~\bibnamefont{Hlozek}},
  \bibinfo{author}{\bibfnamefont{D.~J.~E.} \bibnamefont{Marsh}},
  \bibnamefont{and} \bibinfo{author}{\bibfnamefont{D.}~\bibnamefont{Grin}},
  \bibinfo{journal}{Mon. Not. Roy. Astron. Soc.}
  \textbf{\bibinfo{volume}{476}}, \bibinfo{pages}{3063} (\bibinfo{year}{2018}),
  \eprint{1708.05681}.

\bibitem[{\citenamefont{Hou et~al.}(2013)\citenamefont{Hou, Keisler, Knox,
  Millea, and Reichardt}}]{Hou:2011ec}
\bibinfo{author}{\bibfnamefont{Z.}~\bibnamefont{Hou}},
  \bibinfo{author}{\bibfnamefont{R.}~\bibnamefont{Keisler}},
  \bibinfo{author}{\bibfnamefont{L.}~\bibnamefont{Knox}},
  \bibinfo{author}{\bibfnamefont{M.}~\bibnamefont{Millea}}, \bibnamefont{and}
  \bibinfo{author}{\bibfnamefont{C.}~\bibnamefont{Reichardt}},
  \bibinfo{journal}{Phys. Rev.} \textbf{\bibinfo{volume}{D87}},
  \bibinfo{pages}{083008} (\bibinfo{year}{2013}), \eprint{1104.2333}.

\bibitem[{\citenamefont{Baumann et~al.}(2016)\citenamefont{Baumann, Green,
  Meyers, and Wallisch}}]{Baumann:2015rya}
\bibinfo{author}{\bibfnamefont{D.}~\bibnamefont{Baumann}},
  \bibinfo{author}{\bibfnamefont{D.}~\bibnamefont{Green}},
  \bibinfo{author}{\bibfnamefont{J.}~\bibnamefont{Meyers}}, \bibnamefont{and}
  \bibinfo{author}{\bibfnamefont{B.}~\bibnamefont{Wallisch}},
  \bibinfo{journal}{JCAP} \textbf{\bibinfo{volume}{1601}}, \bibinfo{pages}{007}
  (\bibinfo{year}{2016}), \eprint{1508.06342}.

\bibitem[{\citenamefont{Smith et~al.}(2012)\citenamefont{Smith, Das, and
  Zahn}}]{Smith:2011es}
\bibinfo{author}{\bibfnamefont{T.~L.} \bibnamefont{Smith}},
  \bibinfo{author}{\bibfnamefont{S.}~\bibnamefont{Das}}, \bibnamefont{and}
  \bibinfo{author}{\bibfnamefont{O.}~\bibnamefont{Zahn}},
  \bibinfo{journal}{Phys. Rev.} \textbf{\bibinfo{volume}{D85}},
  \bibinfo{pages}{023001} (\bibinfo{year}{2012}), \eprint{1105.3246}.

\bibitem[{\citenamefont{Sendra and Smith}(2012)}]{Sendra:2012wh}
\bibinfo{author}{\bibfnamefont{I.}~\bibnamefont{Sendra}} \bibnamefont{and}
  \bibinfo{author}{\bibfnamefont{T.~L.} \bibnamefont{Smith}},
  \bibinfo{journal}{Phys. Rev.} \textbf{\bibinfo{volume}{D85}},
  \bibinfo{pages}{123002} (\bibinfo{year}{2012}), \eprint{1203.4232}.

\bibitem[{\citenamefont{Audren et~al.}(2015)}]{Audren:2014lsa}
\bibinfo{author}{\bibfnamefont{B.}~\bibnamefont{Audren}} \bibnamefont{et~al.},
  \bibinfo{journal}{JCAP} \textbf{\bibinfo{volume}{1503}}, \bibinfo{pages}{036}
  (\bibinfo{year}{2015}), \eprint{1412.5948}.

\bibitem[{\citenamefont{Follin et~al.}(2015)\citenamefont{Follin, Knox, Millea,
  and Pan}}]{Follin:2015hya}
\bibinfo{author}{\bibfnamefont{B.}~\bibnamefont{Follin}},
  \bibinfo{author}{\bibfnamefont{L.}~\bibnamefont{Knox}},
  \bibinfo{author}{\bibfnamefont{M.}~\bibnamefont{Millea}}, \bibnamefont{and}
  \bibinfo{author}{\bibfnamefont{Z.}~\bibnamefont{Pan}},
  \bibinfo{journal}{Phys. Rev. Lett.} \textbf{\bibinfo{volume}{115}},
  \bibinfo{pages}{091301} (\bibinfo{year}{2015}), \eprint{1503.07863}.

\bibitem[{\citenamefont{Sellentin and Durrer}(2015)}]{Sellentin:2014gaa}
\bibinfo{author}{\bibfnamefont{E.}~\bibnamefont{Sellentin}} \bibnamefont{and}
  \bibinfo{author}{\bibfnamefont{R.}~\bibnamefont{Durrer}},
  \bibinfo{journal}{Phys. Rev.} \textbf{\bibinfo{volume}{D92}},
  \bibinfo{pages}{063012} (\bibinfo{year}{2015}), \eprint{1412.6427}.

\bibitem[{\citenamefont{Alam et~al.}(2017)}]{Alam:2016hwk}
\bibinfo{author}{\bibfnamefont{S.}~\bibnamefont{Alam}} \bibnamefont{et~al.}
  (\bibinfo{collaboration}{BOSS}), \bibinfo{journal}{Mon. Not. Roy. Astron.
  Soc.} \textbf{\bibinfo{volume}{470}}, \bibinfo{pages}{2617}
  (\bibinfo{year}{2017}), \eprint{1607.03155}.

\bibitem[{\citenamefont{Beutler et~al.}(2011)\citenamefont{Beutler, Blake,
  Colless, Jones, Staveley-Smith, Campbell, Parker, Saunders, and
  Watson}}]{Beutler:2011hx}
\bibinfo{author}{\bibfnamefont{F.}~\bibnamefont{Beutler}},
  \bibinfo{author}{\bibfnamefont{C.}~\bibnamefont{Blake}},
  \bibinfo{author}{\bibfnamefont{M.}~\bibnamefont{Colless}},
  \bibinfo{author}{\bibfnamefont{D.~H.} \bibnamefont{Jones}},
  \bibinfo{author}{\bibfnamefont{L.}~\bibnamefont{Staveley-Smith}},
  \bibinfo{author}{\bibfnamefont{L.}~\bibnamefont{Campbell}},
  \bibinfo{author}{\bibfnamefont{Q.}~\bibnamefont{Parker}},
  \bibinfo{author}{\bibfnamefont{W.}~\bibnamefont{Saunders}}, \bibnamefont{and}
  \bibinfo{author}{\bibfnamefont{F.}~\bibnamefont{Watson}},
  \bibinfo{journal}{Mon. Not. Roy. Astron. Soc.}
  \textbf{\bibinfo{volume}{416}}, \bibinfo{pages}{3017} (\bibinfo{year}{2011}),
  \eprint{1106.3366}.

\bibitem[{\citenamefont{Ross et~al.}(2015)\citenamefont{Ross, Samushia,
  Howlett, Percival, Burden, and Manera}}]{Ross:2014qpa}
\bibinfo{author}{\bibfnamefont{A.~J.} \bibnamefont{Ross}},
  \bibinfo{author}{\bibfnamefont{L.}~\bibnamefont{Samushia}},
  \bibinfo{author}{\bibfnamefont{C.}~\bibnamefont{Howlett}},
  \bibinfo{author}{\bibfnamefont{W.~J.} \bibnamefont{Percival}},
  \bibinfo{author}{\bibfnamefont{A.}~\bibnamefont{Burden}}, \bibnamefont{and}
  \bibinfo{author}{\bibfnamefont{M.}~\bibnamefont{Manera}},
  \bibinfo{journal}{Mon. Not. Roy. Astron. Soc.}
  \textbf{\bibinfo{volume}{449}}, \bibinfo{pages}{835} (\bibinfo{year}{2015}),
  \eprint{1409.3242}.

\bibitem[{\citenamefont{Lewis}(2013)}]{Lewis:2013hha}
\bibinfo{author}{\bibfnamefont{A.}~\bibnamefont{Lewis}},
  \bibinfo{journal}{Phys. Rev.} \textbf{\bibinfo{volume}{D87}},
  \bibinfo{pages}{103529} (\bibinfo{year}{2013}), \eprint{1304.4473}.

\bibitem[{\citenamefont{Gelman and Rubin}(1992)}]{Gelman:1992zz}
\bibinfo{author}{\bibfnamefont{A.}~\bibnamefont{Gelman}} \bibnamefont{and}
  \bibinfo{author}{\bibfnamefont{D.~B.} \bibnamefont{Rubin}},
  \bibinfo{journal}{Statist. Sci.} \textbf{\bibinfo{volume}{7}},
  \bibinfo{pages}{457} (\bibinfo{year}{1992}).

\bibitem[{\citenamefont{Frieman et~al.}(1995)\citenamefont{Frieman, Hill,
  Stebbins, and Waga}}]{Frieman:1995pm}
\bibinfo{author}{\bibfnamefont{J.~A.} \bibnamefont{Frieman}},
  \bibinfo{author}{\bibfnamefont{C.~T.} \bibnamefont{Hill}},
  \bibinfo{author}{\bibfnamefont{A.}~\bibnamefont{Stebbins}}, \bibnamefont{and}
  \bibinfo{author}{\bibfnamefont{I.}~\bibnamefont{Waga}},
  \bibinfo{journal}{Phys. Rev. Lett.} \textbf{\bibinfo{volume}{75}},
  \bibinfo{pages}{2077} (\bibinfo{year}{1995}), \eprint{astro-ph/9505060}.

\bibitem[{\citenamefont{Amendola and Barbieri}(2006)}]{Amendola:2005ad}
\bibinfo{author}{\bibfnamefont{L.}~\bibnamefont{Amendola}} \bibnamefont{and}
  \bibinfo{author}{\bibfnamefont{R.}~\bibnamefont{Barbieri}},
  \bibinfo{journal}{Phys. Lett.} \textbf{\bibinfo{volume}{B642}},
  \bibinfo{pages}{192} (\bibinfo{year}{2006}), \eprint{hep-ph/0509257}.

\bibitem[{\citenamefont{Sarkar et~al.}(2015)\citenamefont{Sarkar, Das, and
  Sethi}}]{Sarkar:2014bca}
\bibinfo{author}{\bibfnamefont{A.}~\bibnamefont{Sarkar}},
  \bibinfo{author}{\bibfnamefont{S.}~\bibnamefont{Das}}, \bibnamefont{and}
  \bibinfo{author}{\bibfnamefont{S.~K.} \bibnamefont{Sethi}},
  \bibinfo{journal}{JCAP} \textbf{\bibinfo{volume}{1503}}, \bibinfo{pages}{004}
  (\bibinfo{year}{2015}), \eprint{1410.7129}.

\bibitem[{\citenamefont{Carroll and Kaplinghat}(2002)}]{Carroll:2001bv}
\bibinfo{author}{\bibfnamefont{S.~M.} \bibnamefont{Carroll}} \bibnamefont{and}
  \bibinfo{author}{\bibfnamefont{M.}~\bibnamefont{Kaplinghat}},
  \bibinfo{journal}{Phys. Rev.} \textbf{\bibinfo{volume}{D65}},
  \bibinfo{pages}{063507} (\bibinfo{year}{2002}), \eprint{astro-ph/0108002}.

\bibitem[{\citenamefont{Abazajian et~al.}(2016)}]{Abazajian:2016yjj}
\bibinfo{author}{\bibfnamefont{K.~N.} \bibnamefont{Abazajian}}
  \bibnamefont{et~al.} (\bibinfo{collaboration}{CMB-S4})
  (\bibinfo{year}{2016}), \eprint{1610.02743}.

\bibitem[{\citenamefont{Freedman}(2017)}]{Freedman:2017yms}
\bibinfo{author}{\bibfnamefont{W.~L.} \bibnamefont{Freedman}},
  \bibinfo{journal}{Nat. Astron.} \textbf{\bibinfo{volume}{1}},
  \bibinfo{pages}{0121} (\bibinfo{year}{2017}), \eprint{1706.02739}.

\bibitem[{\citenamefont{Bernal et~al.}(2016)\citenamefont{Bernal, Verde, and
  Riess}}]{Bernal:2016gxb}
\bibinfo{author}{\bibfnamefont{J.~L.} \bibnamefont{Bernal}},
  \bibinfo{author}{\bibfnamefont{L.}~\bibnamefont{Verde}}, \bibnamefont{and}
  \bibinfo{author}{\bibfnamefont{A.~G.} \bibnamefont{Riess}},
  \bibinfo{journal}{JCAP} \textbf{\bibinfo{volume}{1610}}, \bibinfo{pages}{019}
  (\bibinfo{year}{2016}), \eprint{1607.05617}.

\bibitem[{\citenamefont{Aubourg et~al.}(2015)}]{Aubourg:2014yra}
\bibinfo{author}{\bibfnamefont{E.}~\bibnamefont{Aubourg}} \bibnamefont{et~al.},
  \bibinfo{journal}{Phys. Rev.} \textbf{\bibinfo{volume}{D92}},
  \bibinfo{pages}{123516} (\bibinfo{year}{2015}), \eprint{1411.1074}.

\bibitem[{\citenamefont{Dvorkin et~al.}(2014)\citenamefont{Dvorkin, Wyman,
  Rudd, and Hu}}]{Dvorkin:2014lea}
\bibinfo{author}{\bibfnamefont{C.}~\bibnamefont{Dvorkin}},
  \bibinfo{author}{\bibfnamefont{M.}~\bibnamefont{Wyman}},
  \bibinfo{author}{\bibfnamefont{D.~H.} \bibnamefont{Rudd}}, \bibnamefont{and}
  \bibinfo{author}{\bibfnamefont{W.}~\bibnamefont{Hu}}, \bibinfo{journal}{Phys.
  Rev.} \textbf{\bibinfo{volume}{D90}}, \bibinfo{pages}{083503}
  (\bibinfo{year}{2014}), \eprint{1403.8049}.

\bibitem[{\citenamefont{Wyman et~al.}(2014)\citenamefont{Wyman, Rudd,
  Vanderveld, and Hu}}]{Wyman:2013lza}
\bibinfo{author}{\bibfnamefont{M.}~\bibnamefont{Wyman}},
  \bibinfo{author}{\bibfnamefont{D.~H.} \bibnamefont{Rudd}},
  \bibinfo{author}{\bibfnamefont{R.~A.} \bibnamefont{Vanderveld}},
  \bibnamefont{and} \bibinfo{author}{\bibfnamefont{W.}~\bibnamefont{Hu}},
  \bibinfo{journal}{Phys. Rev. Lett.} \textbf{\bibinfo{volume}{112}},
  \bibinfo{pages}{051302} (\bibinfo{year}{2014}), \eprint{1307.7715}.

\bibitem[{\citenamefont{Leistedt et~al.}(2014)\citenamefont{Leistedt, Peiris,
  and Verde}}]{Leistedt:2014sia}
\bibinfo{author}{\bibfnamefont{B.}~\bibnamefont{Leistedt}},
  \bibinfo{author}{\bibfnamefont{H.~V.} \bibnamefont{Peiris}},
  \bibnamefont{and} \bibinfo{author}{\bibfnamefont{L.}~\bibnamefont{Verde}},
  \bibinfo{journal}{Phys. Rev. Lett.} \textbf{\bibinfo{volume}{113}},
  \bibinfo{pages}{041301} (\bibinfo{year}{2014}), \eprint{1404.5950}.

\bibitem[{\citenamefont{Ade et~al.}(2016{\natexlab{b}})}]{Ade:2015rim}
\bibinfo{author}{\bibfnamefont{P.~A.~R.} \bibnamefont{Ade}}
  \bibnamefont{et~al.} (\bibinfo{collaboration}{Planck}),
  \bibinfo{journal}{Astron. Astrophys.} \textbf{\bibinfo{volume}{594}},
  \bibinfo{pages}{A14} (\bibinfo{year}{2016}{\natexlab{b}}),
  \eprint{1502.01590}.

\bibitem[{\citenamefont{Di~Valentino et~al.}(2016)\citenamefont{Di~Valentino,
  Melchiorri, and Silk}}]{DiValentino:2016hlg}
\bibinfo{author}{\bibfnamefont{E.}~\bibnamefont{Di~Valentino}},
  \bibinfo{author}{\bibfnamefont{A.}~\bibnamefont{Melchiorri}},
  \bibnamefont{and} \bibinfo{author}{\bibfnamefont{J.}~\bibnamefont{Silk}},
  \bibinfo{journal}{Phys. Lett.} \textbf{\bibinfo{volume}{B761}},
  \bibinfo{pages}{242} (\bibinfo{year}{2016}), \eprint{1606.00634}.

\bibitem[{\citenamefont{Di~Valentino et~al.}(2018)\citenamefont{Di~Valentino,
  Linder, and Melchiorri}}]{DiValentino:2017rcr}
\bibinfo{author}{\bibfnamefont{E.}~\bibnamefont{Di~Valentino}},
  \bibinfo{author}{\bibfnamefont{E.~V.} \bibnamefont{Linder}},
  \bibnamefont{and}
  \bibinfo{author}{\bibfnamefont{A.}~\bibnamefont{Melchiorri}},
  \bibinfo{journal}{Phys. Rev.} \textbf{\bibinfo{volume}{D97}},
  \bibinfo{pages}{043528} (\bibinfo{year}{2018}), \eprint{1710.02153}.

\bibitem[{\citenamefont{Di~Valentino et~al.}(2017)\citenamefont{Di~Valentino,
  Melchiorri, and Mena}}]{DiValentino:2017iww}
\bibinfo{author}{\bibfnamefont{E.}~\bibnamefont{Di~Valentino}},
  \bibinfo{author}{\bibfnamefont{A.}~\bibnamefont{Melchiorri}},
  \bibnamefont{and} \bibinfo{author}{\bibfnamefont{O.}~\bibnamefont{Mena}},
  \bibinfo{journal}{Phys. Rev.} \textbf{\bibinfo{volume}{D96}},
  \bibinfo{pages}{043503} (\bibinfo{year}{2017}), \eprint{1704.08342}.

\bibitem[{\citenamefont{Poulin et~al.}(2018)\citenamefont{Poulin, Boddy, Bird,
  and Kamionkowski}}]{Poulin:2018zxs}
\bibinfo{author}{\bibfnamefont{V.}~\bibnamefont{Poulin}},
  \bibinfo{author}{\bibfnamefont{K.~K.} \bibnamefont{Boddy}},
  \bibinfo{author}{\bibfnamefont{S.}~\bibnamefont{Bird}}, \bibnamefont{and}
  \bibinfo{author}{\bibfnamefont{M.}~\bibnamefont{Kamionkowski}}
  (\bibinfo{year}{2018}), \eprint{1803.02474}.

\bibitem[{\citenamefont{Ewall-Wice et~al.}(2018)\citenamefont{Ewall-Wice,
  Chang, Lazio, Dore, Seiffert, and Monsalve}}]{Ewall-Wice:2018bzf}
\bibinfo{author}{\bibfnamefont{A.}~\bibnamefont{Ewall-Wice}},
  \bibinfo{author}{\bibfnamefont{T.~C.} \bibnamefont{Chang}},
  \bibinfo{author}{\bibfnamefont{J.}~\bibnamefont{Lazio}},
  \bibinfo{author}{\bibfnamefont{O.}~\bibnamefont{Dore}},
  \bibinfo{author}{\bibfnamefont{M.}~\bibnamefont{Seiffert}}, \bibnamefont{and}
  \bibinfo{author}{\bibfnamefont{R.~A.} \bibnamefont{Monsalve}}
  (\bibinfo{year}{2018}), \eprint{1803.01815}.

\bibitem[{\citenamefont{Mirocha and Furlanetto}(2018)}]{Mirocha:2018cih}
\bibinfo{author}{\bibfnamefont{J.}~\bibnamefont{Mirocha}} \bibnamefont{and}
  \bibinfo{author}{\bibfnamefont{S.~R.} \bibnamefont{Furlanetto}}
  (\bibinfo{year}{2018}), \eprint{1803.03272}.

\bibitem[{\citenamefont{Pospelov et~al.}(2018)\citenamefont{Pospelov, Pradler,
  Ruderman, and Urbano}}]{Pospelov:2018kdh}
\bibinfo{author}{\bibfnamefont{M.}~\bibnamefont{Pospelov}},
  \bibinfo{author}{\bibfnamefont{J.}~\bibnamefont{Pradler}},
  \bibinfo{author}{\bibfnamefont{J.~T.} \bibnamefont{Ruderman}},
  \bibnamefont{and} \bibinfo{author}{\bibfnamefont{A.}~\bibnamefont{Urbano}}
  (\bibinfo{year}{2018}), \eprint{1803.07048}.

\bibitem[{\citenamefont{Barkana}(2018)}]{Barkana:2018lgd}
\bibinfo{author}{\bibfnamefont{R.}~\bibnamefont{Barkana}},
  \bibinfo{journal}{Nature} \textbf{\bibinfo{volume}{555}}, \bibinfo{pages}{71}
  (\bibinfo{year}{2018}), \eprint{1803.06698}.

\bibitem[{\citenamefont{Costa et~al.}(2018)\citenamefont{Costa, Landim, Wang,
  and Abdalla}}]{Costa:2018aoy}
\bibinfo{author}{\bibfnamefont{A.~A.} \bibnamefont{Costa}},
  \bibinfo{author}{\bibfnamefont{R.~C.~G.} \bibnamefont{Landim}},
  \bibinfo{author}{\bibfnamefont{B.}~\bibnamefont{Wang}}, \bibnamefont{and}
  \bibinfo{author}{\bibfnamefont{E.}~\bibnamefont{Abdalla}}
  (\bibinfo{year}{2018}), \eprint{1803.06944}.

\bibitem[{\citenamefont{Seager et~al.}(2000)\citenamefont{Seager, Sasselov, and
  Scott}}]{Seager:1999km}
\bibinfo{author}{\bibfnamefont{S.}~\bibnamefont{Seager}},
  \bibinfo{author}{\bibfnamefont{D.~D.} \bibnamefont{Sasselov}},
  \bibnamefont{and} \bibinfo{author}{\bibfnamefont{D.}~\bibnamefont{Scott}},
  \bibinfo{journal}{Astrophys. J. Suppl.} \textbf{\bibinfo{volume}{128}},
  \bibinfo{pages}{407} (\bibinfo{year}{2000}), \eprint{astro-ph/9912182}.

\bibitem[{\citenamefont{Seager et~al.}(1999)\citenamefont{Seager, Sasselov, and
  Scott}}]{Seager:1999bc}
\bibinfo{author}{\bibfnamefont{S.}~\bibnamefont{Seager}},
  \bibinfo{author}{\bibfnamefont{D.~D.} \bibnamefont{Sasselov}},
  \bibnamefont{and} \bibinfo{author}{\bibfnamefont{D.}~\bibnamefont{Scott}},
  \bibinfo{journal}{Astrophys. J.} \textbf{\bibinfo{volume}{523}},
  \bibinfo{pages}{L1} (\bibinfo{year}{1999}), \eprint{astro-ph/9909275}.

\bibitem[{\citenamefont{Ali-Haimoud and Hirata}(2011)}]{AliHaimoud:2010dx}
\bibinfo{author}{\bibfnamefont{Y.}~\bibnamefont{Ali-Haimoud}} \bibnamefont{and}
  \bibinfo{author}{\bibfnamefont{C.~M.} \bibnamefont{Hirata}},
  \bibinfo{journal}{Phys. Rev.} \textbf{\bibinfo{volume}{D83}},
  \bibinfo{pages}{043513} (\bibinfo{year}{2011}), \eprint{1011.3758}.

\bibitem[{\citenamefont{Smith et~al.}()\citenamefont{Smith, Poulin, Grin,
  Hlozek, Marsh, and Scott}}]{SmithXXX}
\bibinfo{author}{\bibfnamefont{T.}~\bibnamefont{Smith}},
  \bibinfo{author}{\bibfnamefont{V.}~\bibnamefont{Poulin}},
  \bibinfo{author}{\bibfnamefont{D.}~\bibnamefont{Grin}},
  \bibinfo{author}{\bibfnamefont{R.}~\bibnamefont{Hlozek}},
  \bibinfo{author}{\bibfnamefont{D.}~\bibnamefont{Marsh}}, \bibnamefont{and}
  \bibinfo{author}{\bibfnamefont{M.}~\bibnamefont{Scott}}, \bibinfo{journal}{in
  prep.}  (????).

\bibitem[{\citenamefont{Cookmeyer et~al.}()\citenamefont{Cookmeyer, Grin, and
  Smith}}]{JonnyCAMB}
\bibinfo{author}{\bibfnamefont{J.}~\bibnamefont{Cookmeyer}},
  \bibinfo{author}{\bibfnamefont{D.}~\bibnamefont{Grin}}, \bibnamefont{and}
  \bibinfo{author}{\bibfnamefont{T.}~\bibnamefont{Smith}}, \bibinfo{journal}{in
  prep.}  (????).

\bibitem[{\citenamefont{Owen et~al.}()\citenamefont{Owen, Poulin, Smith, Grin,
  and Kamionkowski}}]{AxiCLASS}
\bibinfo{author}{\bibfnamefont{C.}~\bibnamefont{Owen}},
  \bibinfo{author}{\bibfnamefont{V.}~\bibnamefont{Poulin}},
  \bibinfo{author}{\bibfnamefont{T.}~\bibnamefont{Smith}},
  \bibinfo{author}{\bibfnamefont{D.}~\bibnamefont{Grin}}, \bibnamefont{and}
  \bibinfo{author}{\bibfnamefont{M.}~\bibnamefont{Kamionkowski}},
  \bibinfo{journal}{in prep.}  (????).

\bibitem[{\citenamefont{Hu}(2004)}]{Hu:2004xd}
\bibinfo{author}{\bibfnamefont{W.}~\bibnamefont{Hu}}, in
  \emph{\bibinfo{booktitle}{{Astroparticle physics and cosmology. Proceedings:
  Summer School, Trieste, Italy, Jun 17-Jul 5 2002}}} (\bibinfo{year}{2004}),
  \eprint{astro-ph/0402060}.

\end{thebibliography}

\end{appendix}

\end{document}